\documentclass[12pt]{article}
\usepackage{amsmath}
\usepackage{amssymb} % e.g. \gtrsim
\usepackage{amsthm} % e.g. \gtrsim
\usepackage{bbm} % BlackBoeard letters
\usepackage[small]{caption2} % font of captions is different from text
\usepackage{fleqn} % equations are not centered
\usepackage{graphicx} % including PostScript
\usepackage{mathrsfs}	% e.g. nice Lagrange-L with \mathscr{L}
\usepackage{paralist}
\usepackage{pifont}
\usepackage{picins} 
\usepackage{pstricks} 
\usepackage{pst-node} 
\usepackage{pst-plot} 
\usepackage{pst-slpe}
\usepackage{pst-blur}
\usepackage[small,loose]{subfigure}  % subfigures with (a), (b) etc., ... and own caption
\makeatletter
\advance \headheight by 3.0truept       % for 12pt mandatory...
\setcaptionwidth{0.75\textwidth}
\definecolor{darkgreen}{rgb}{0.0, 0.6, 0.2}
\newrgbcolor{darkgreen}{0.0 0.6 0.2}
\definecolor{purple}{rgb}{0.75, 0.2, 0.75}
\newrgbcolor{purple}{0.75 0.2 0.75}
\newgray{gray90}{0.9}
\newgray{gray80}{0.8}
\newgray{gray70}{0.7}

\newcommand{\QuadraticPillow}[9]{
\begin{pspicture}(-3,-0.8)(7,5)
	\pscurve[linewidth=0.5mm](0,0)(2,0.2)(4,0)
	\pscurve[linewidth=0.5mm](0.8,0.28)(2,0.4)(3.2,0.28)
	\pscurve[linewidth=0.5mm](1.2,0.5)(2,0.56)(2.8,0.5)
	\pscurve[linewidth=0.5mm](0,4)(2,3.8)(4,4)
	\pscurve[linewidth=0.5mm](0.8,3.72)(2,3.6)(3.2,3.72)
	\pscurve[linewidth=0.5mm](1.2,3.5)(2,3.44)(2.8,3.5)
	\pscurve[linewidth=0.5mm](0,0)(0.3,2)(0,4)
	\pscurve[linewidth=0.5mm](0.36,0.8)(0.48,2)(0.36,3.2)
	\pscurve[linewidth=0.5mm](0.6,1.2)(0.68,2)(0.6,2.8)
	\pscurve[linewidth=0.5mm](4,0)(3.7,2)(4,4)
	\pscurve[linewidth=0.5mm](3.64,0.8)(3.52,2)(3.64,3.2)
	\pscurve[linewidth=0.5mm](3.4,1.2)(3.32,2)(3.4,2.8)
	\psdot*[dotstyle=o,dotsize=3mm](0,0)
 	\psdot*[dotstyle=square*,dotsize=2mm](0,0)
	\rput[rt](1.6,-0.2){#1}
	\rput[rb](-0.4,0.2){#2}
	\psdot*[dotstyle=o,dotsize=3mm](4,0)
	\psdot*[dotstyle=square*,dotsize=2mm](4,0)
	\rput[lt](2.4,-0.2){#3}
	\rput[lb](4.4,0.2){#4}	
	\psdot*[dotstyle=o,dotsize=3mm](4,4)
	\psdot*[dotstyle=square*,dotsize=2mm](4,4)
	\rput[rb](1.6,4.2){#5}
	\rput[rt](-0.4,3.8){#6}	
	\psdot*[dotstyle=o,dotsize=3mm](0,4)
	\psdot*[dotstyle=square*,dotsize=2mm](0,4)
	\rput[lb](2.4,4.2){#7}
	\rput[lt](4.2,3.8){#8}	
	\rput[c](2,2){#9}
\end{pspicture}
}

\newcommand{\CornerA}[6]{\begin{pspicture}(-1,-1)(2.0,2.8)
	\pscurve[linewidth=0.5mm](0,0)(1,0.15)(2,0.2)
	\pscurve[linewidth=0.5mm](0.8,0.28)(1.4,0.35)(2,0.4)
	\pscurve[linewidth=0.5mm](1.2,0.5)(1.6,0.54)(2,0.56)
	\pscurve[linewidth=0.5mm](0,0)(0.15,1)(0.2,2)
	\pscurve[linewidth=0.5mm](0.28,0.8)(0.35,1.4)(0.4,2)
	\pscurve[linewidth=0.5mm](0.5,1.2)(0.54,1.6)(0.56,2)
	\psline[linewidth=0.5mm](2,0.2)(2,0.9)(1.7,0.9)
	\psline[linewidth=0.5mm](2,2)(2,1.3)(1.7,1.3)
	\psarc[linewidth=0.5mm](1.5,1.1){0.28}{45}{315}
	\psline[linewidth=0.5mm](0.2,2)(0.9,2)(0.9,2.3)
	\psline[linewidth=0.5mm](2,2)(1.3,2)(1.3,2.3)
	\psarc[linewidth=0.5mm](1.1,2.5){0.28}{-45}{225}
	\psdot*[dotstyle=o,dotsize=3mm](0,0)
 	\psdot*[dotstyle=*,dotsize=2mm](0,0)
	\rput[#2]{#1}(0,-#6){#4}
	\rput[#3]{#1}(-#6,0){#5}
\end{pspicture}
}

\newcommand{\CenterObject}[1]{\ensuremath{\vcenter{\hbox{#1}}}}
 
\newcommand{\I}{\mathrm{i}}

\newcommand{\E}[1]{\ensuremath{\mathrm{E}_{#1}}} % e.g. \E{8}
\newcommand{\G}[1]{\ensuremath{\mathrm{G}_{#1}}}
\newcommand{\SO}[1]{\ensuremath{\mathrm{SO}(#1)}}
\newcommand{\SU}[1]{\ensuremath{\mathrm{SU}(#1)}}
\newcommand{\U}[1]{\ensuremath{\mathrm{U}(#1)}}
\newcommand{\Z}[1]{\ensuremath{\mathbbm{Z}_{#1}}} % Z_N ->\Z{N}
\DeclareMathOperator{\diag}{diag}
\DeclareMathOperator{\ad}{ad}
\DeclareMathOperator{\re}{Re}

\begin{document}
\title{\vspace*{-3cm}{\normalsize\hfill TUM-HEP 07/678}\\[2.5cm]
\textbf{Notes on Local Grand Unification\footnote{Based
on lectures given at the Summer Institute 2007, Fuji-Yoshida, Japan. The slides can
be found at \texttt{http://wwwhep.s.kanazawa-u.ac.jp/SI2007/slides/ratz.pdf}.} }}
\author{Michael Ratz\\
{\it\normalsize Physik Department T30, Technische Universit\"at M\"unchen,}\\[-0.05cm]
{\it\normalsize James-Franck-Strasse, 85748 Garching, Germany}}
\date{November 10, 2007}
\maketitle
\begin{abstract}
Grand unified models in four dimensions typically suffer from the
doublet-triplet splitting problem. This obstacle can be overcome in
higher-dimensional settings, where a non-trivial gauge group topography can
explain the simultaneous appearance of complete standard model generations in
the form of $\boldsymbol{16}$-plets of SO(10) and the Higgs fields as split
multiplets. In these notes, the emerging scheme of `local grand unification' and
its realization in the context of orbifold compactifications of the heterotic
string are reviewed.
\end{abstract}
\noindent

\clearpage
\section{Introduction}

The standard model (SM) is remarkably successful in explaining observations. It
is based on the gauge group 
\begin{equation}
 G_\mathrm{SM}~=~\SU3_\mathrm{C}\times\SU2_\mathrm{L}\times\U1_Y\;,
\end{equation}
and has three generations of matter transforming as
\begin{equation}\label{eq:SMgeneration}
        (\boldsymbol{3},\boldsymbol{2})_{1/6}+
        (\overline{\boldsymbol{3}},\boldsymbol{1})_{-2/3}+
        (\overline{\boldsymbol{3}},\boldsymbol{1})_{1/3}+
        (\boldsymbol{1},\boldsymbol{2})_{-1/2}+
        (\boldsymbol{1},\boldsymbol{1})_{1}\;.
\end{equation}
It further contains the Higgs field which transforms as
$(\boldsymbol{1},\boldsymbol{2})_{1/2}$.  The evidence for neutrino masses
strongly supports the existence of right-handed neutrinos, which are singlets
under $G_\mathrm{SM}$. One of the features of SM matter that points towards a
deeper underlying structure is charge quantization, i.e.\ the fact that 
hypercharges of different multiplets are integer multiples of a common unit. The
most compelling explanation of this fact arises in schemes where
hypercharge is embedded in a non-Abelian group factor \cite{Pati:1974yy}, in
particular in grand unified theories (GUTs) \cite{Georgi:1974sy}. 
In the context of an \SU5 GUT, the five irreducible representations of
\eqref{eq:SMgeneration} can be combined into two \SU5 representations
\cite{Georgi:1974sy},
\begin{subequations}\label{eq:SU5matter}
\begin{eqnarray} 
 \boldsymbol{10}
 & = &  (\boldsymbol{3},\boldsymbol{2})_{1/6}\oplus
        (\overline{\boldsymbol{3}},\boldsymbol{1})_{-2/3}\oplus
        (\boldsymbol{1},\boldsymbol{1})_{1}\;,\\
 \overline{\boldsymbol{5}}		
 & = &
        (\overline{\boldsymbol{3}},\boldsymbol{1})_{1/3}\oplus
        (\boldsymbol{1},\boldsymbol{2})_{-1/2}\;.
\end{eqnarray}
\end{subequations}
That is, one generation of standard model fermions (without the right-handed
neutrino) can be combined into two irreducible representations of \SU5.
Hypercharge is in the Cartan basis of \SU5 but not in the Cartan bases of
$\SU3_\mathrm{C}$ or $\SU2_\mathrm{L}$.

Even more impressive than the \SU5 relations \eqref{eq:SU5matter} 
is the fact that a
$\boldsymbol{16}$-plet of \SO{10} comprises one full SM generation including the
right-handed neutrino \cite{Georgi:1975qb,Fritzsch:1974nn},
\begin{equation}
\boldsymbol{16}~=~
(\boldsymbol{3}, \boldsymbol{2})_{1/6} \oplus
(\boldsymbol{\overline{3}}, \boldsymbol{1})_{-2/3}\oplus
(\boldsymbol{\overline{3}}, \boldsymbol{1})_{1/3}\oplus
(\boldsymbol{1}, \boldsymbol{2})_{-1/2} \oplus
(\boldsymbol{1}, \boldsymbol{1})_{1}\oplus
(\boldsymbol{1}, \boldsymbol{1})_{0}
 \;.
\end{equation}

While charge quantization can also be attributed to the requirement of anomaly cancellation, the scheme
of GUTs receives further support from gauge coupling unification. It
is well known that the gauge couplings unify in the minimal supersymmetric
extensions of the standard model (MSSM) \cite{Amaldi:1991cn} at the scale 
\begin{equation}
 M_\mathrm{GUT}~\simeq~2\cdot 10^{16}\,\mathrm{GeV}\;.
\end{equation}
In these notes, we will take this observation seriously and from now on consider
theories with low-energy supersymmetry that are consistent with MSSM gauge
coupling unification.

Unfortunately, the SM Higgs sector casts some shadow on the standard picture of
GUTs. The smallest \SO{10} representation containing the MSSM Higgs
doublets (or the SM Higgs) is the $\boldsymbol{10}$-plet, which decomposes as 
\begin{equation}
 \boldsymbol{10}~=~
 (\boldsymbol{1},\boldsymbol{2})_{1/2}
 \oplus
 (\boldsymbol{1},\boldsymbol{2})_{-1/2}
 \oplus
 (\boldsymbol{3},\boldsymbol{1})_{-1/3}
 \oplus
 (\overline{\boldsymbol{3}},\boldsymbol{1})_{1/3}\;.
\end{equation}
\piccaption{Proton decay via triplet exchange.}
\parpic(6cm,3cm)[r]{
\CenterObject{\includegraphics{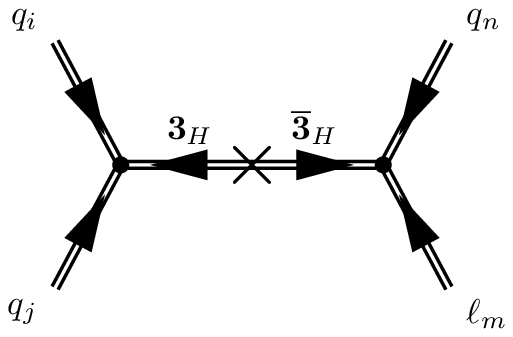}}
\label{fig:TripletExchange}
}
While \SU2 doublets are required for electroweak symmetry breaking, there are
strong experimental constraints on the triplet masses. They come from effective
dimension 5 operators \cite{Sakai:1981pk,Weinberg:1981wj,Dimopoulos:1981dw} that arise from
integrating out the triplets and lead to proton decay (cf.\
\cite{Dermisek:2000hr}); see figure~\ref{fig:TripletExchange}. This leads, first
of all, to the so-called doublet-triplet splitting problem, which can be phrased
as:
\picskip{0}
\begin{quote}
 ``Why does matter appear in complete GUT representations while the Higgs
 multiplets are incomplete (split) ?''
\end{quote}
Note that the problem has two aspects. On the one hand, in the context of
grand unification one might ask why the Higgses are split, on the other hand in a
theory without unification one would like to understand why matter apparently
fits into representations of a larger group.

Although there have been ingenious proposals to solve the problem in the context
of four-dimensional (4D) grand unification (see e.g.\
\cite{Dimopoulos:1981xm,Masiero:1982fe}), it is probably fair to say that none
of these renders the theory as  simple and  beautiful as the original idea of GUTs. 
Besides,
these solutions give a GUT scale mass to the triplets, which might not be enough
to suppress the dimension 5 proton decay operators \cite{Dermisek:2000hr}.

\section{Orbifold GUTs}

Arguably, the most appealing solutions to the doublet-triplet splitting problem
are obtained in higher-dimensional extensions of the standard model. While this
point has been stressed from early on in the string literature (cf.\
\cite{Witten:1985xc,Breit:1985ud}), in these notes we will use the framework of
orbifold GUTs \cite{Kawamura:1999nj,Kawamura:2000ev,Altarelli:2001qj,%
Hall:2001pg,Hebecker:2001wq,Asaka:2001eh,Hall:2001xr} (for a review, see e.g.\
\cite{Quiros:2003gg}) in order to see how this works. Later, in sections 
\ref{sec:StringOrbifolds} and \ref{sec:LGU}, these constructions will be
embedded into string theory.

From now on, let us entertain the possibility that the known four space time
dimensions are amended by extra compact dimensions. That is, we consider
gauge theories, based on the gauge group $G$, on the space
\begin{equation}
 \mathbbm{M}^4\times K\;,
\end{equation}
where $\mathbbm{M}^4$ is the usual Minkowski space and $K$ is compact.
There are some (phenomenological) constraints on the properties of the internal
space $K$. The most important requirement for the subsequent discussion is that
$K$ be such as to give rise to a chiral spectrum in 4D effective field theory
which emerges at scales below the compactification scale. 

\subsection{One-dimensional orbifold GUTs}

\begin{figure}[h]
\centerline{\begin{pspicture}(-4.5,-3.5)(4,3.5)
	\pscircle(0,0){3}
	\psline[linestyle=dashed]{-}(-4,0)(4,0)
	\rput[r](-4.1,0){$\blue P$}
	\psline[linewidth=1mm,linecolor=blue]{<->}(-1.5,-2.4)(-1.5,2.4)
	\psline[linewidth=1mm,linecolor=blue]{<->}(0,-2.9)(-0,2.9)
	\psline[linewidth=1mm,linecolor=blue]{<->}(1.5,-2.4)(1.5,2.4)
	\psarc[linewidth=1mm,linecolor=darkgreen]{-}(0,0){3}{00}{180}
	\psarc[linecolor=darkgreen]{<-}(0,0){3.2}{140}{180}
	\rput[r](-2.8,2){$\darkgreen x_5$}
	\pscircle*[linecolor=purple](-3.02,0){0.15}
	\pscircle*[linecolor=purple](2.98,0){0.15}
\end{pspicture}
}
\caption{$\mathbbm{S}^1/\mathbbm{Z}_2$ orbifold.}
\label{fig:S1overZ2}
\end{figure}
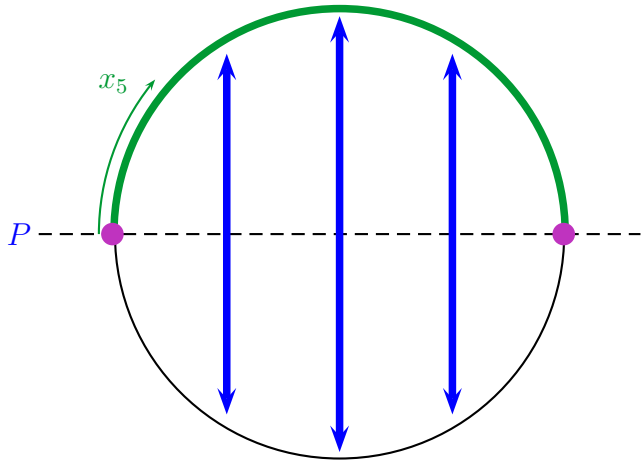

The  simplest setting satisfying this requirement is an
$\mathbbm{S}^1/\mathbbm{Z}_2$ orbifold (figure~\ref{fig:S1overZ2}). An
orbifold, in general, emerges from  dividing a manifold by one of its (non-freely
acting) discrete symmetries. In the $\mathbbm{S}^1/\mathbbm{Z}_2$ case one first
compactifies one extra dimension on a circle and then  identifies points related
by a \Z2 reflection symmetry, which is a symmetry of the circle. The emerging
space, that is the fundamental domain of the orbifold,  is an interval which is
bounded by the orbifold fixed points, i.e.\ the two points that are invariant
under the orbifold action. 

The important point in the context of GUT model building is that one can
associate an automorphism of the gauge
group with  the reflection in the  internal space. 
That is, the orbifold action is
\begin{eqnarray}
 \Theta & : & x_5~\to~\theta\,x_5~=~-x_5\;,\nonumber\\
 & & \boldsymbol{r}~\to~P_{\boldsymbol{r}}\,\boldsymbol{r}\;.
\end{eqnarray}
Here, $\boldsymbol{r}$ is a representation of $G$ and $P_{\boldsymbol{r}}$ is
the representation (matrix) of the automorphism. Since $P$ squares to the
identity in internal space, we will require that
$R_{\boldsymbol{r}}(P)^2=\mathbbm{1}$.\footnote{This requirement might
potentially be relaxed, see \cite{Hebecker:2003jt,vonGersdorff:2007uz}.} 

The most important application of the embedding of the orbifold action in the
gauge degrees of freedom is gauge symmetry breaking by orbifolding. Consider
$\mathbbm{M}^4\times \mathbbm{S}^1/\mathbbm{Z}_2$ and  a state transforming
under the representation $\boldsymbol{r}$. The requirement that the state be
invariant under the orbifold action amounts to
\begin{equation}
 \phi_{\boldsymbol{r}}(x_0,\dots x_3;x_5)
 ~=~P_{\boldsymbol{r}}\,\phi_{\boldsymbol{r}}(x_0,\dots x_3;-x_5)\;.
\end{equation}
In the \Z2 case under consideration this leads to various  ``orbifold
parities''.

Consider, as an example, the orbifold $\mathbbm{S}^1/\mathbbm{Z}_2$ with gauge
group \SU3. In this example, let us focus on  a non-supersymmetric setting. For
the automorphism in the fundamental representation take  
\begin{equation}
 P~=~\diag(-1,-1,1)\;.
\end{equation}
This means that scalar bulk fields transforming as $\boldsymbol{3}$-plets under
\SU3 have to satisfy
\begin{equation}\label{eq:OrbifoldActionOnFields1}
 \phi(x_\mu,-x_5)~=~P\,\phi(x_\mu,x_5)\;.
\end{equation}
This implies the boundary condition
\begin{equation}
 \phi(x,0)~=~P\,\phi(x,0)
\end{equation}
on the interval. Since $x_5=L$ and $x_5=-L$ coincide, one also has
\begin{equation}
 \phi(x,L)~=~P\,\phi(x,L)\;.
\end{equation}
Let us now work with the eigenstates of $P$. 
The field $\phi$ decomposes into two pieces, the eigenstates of $P$ with
eigenvalues $+1$ and $-1$, which we will denote by $\phi_\pm$, respectively.
In this notation, equation \eqref{eq:OrbifoldActionOnFields1} reads 
\begin{equation}
 \phi_\pm(x_\mu,-x_5)
 ~=~P\,\phi_\pm(x_\mu,x_5)~=~\pm\,\phi_\pm(x_\mu,x_5)
 \;.
\end{equation}
Since the fifth dimension is compact, we can expand the eigenstates in terms of
Fourier modes \cite{Kawamura:1999nj},
\begin{subequations}
\begin{eqnarray}
 \phi_{+}(x_\mu,x_5) & = &
 	\sum_{n=0}^\infty \frac{1}{\sqrt{2^{\delta_{n0}}\pi\, R}}\,
		\phi_{+}^{(n)}(x_\mu)\,\cos\left(\frac{n\,x_5}{R}\right)
 \;,\\
 \phi_{-}(x_\mu,x_5) & = &
 	\sum_{n=1}^\infty \frac{1}{\sqrt{\pi\, R}}\,\phi_{-}^{(n)}(x_\mu)\,
		\sin\left(\frac{ n\,x_5}{R}\right)
 \;.
\end{eqnarray}
\end{subequations}
Here $R$ is the radius of the $\mathbbm{S}^1$ and the length of the interval is
therefore $L=R\,\pi$.
In the 4-dimensional effective theory, the Fourier modes can be regarded as 
massive states with masses $n/R$. In particular, only the $\phi_+$ field possesses a
zero-mode. This implies that at energies below $1/R$ one can integrate out
the heavy modes, thus obtaining an effective theory with only the $\phi_+$
zero-modes as dynamical degrees of freedom.

An analogous analysis can be carried out for the gauge fields. Here, the
boundary conditions for the four-dimensional vector fields read
\begin{equation}\label{eq:ProjectionA1}
 A_\mu^a(x_\mu,-x_5)\,\mathsf{T}_a~=~
 A_\mu^a(x_\mu,x_5)\,P\,\mathsf{T}_a\,P^{-1}\;,
\end{equation}
with $\mathsf{T}_a$ denoting the generators. It is straightforward to check that
only the gauge bosons of an $\SU2\times\U1$ subgroup of the original \SU3 have
zero-modes in 4D. For instance, a simple explicit calculation with the standard
\SU3 matrices (see e.g.\ \cite[p.~502]{Peskin:1995ev}) reveals that only the
gauge bosons associated with $\lambda_i/2$ where $i\in\{1,2,3,8\}$ survive the
projection condition \eqref{eq:ProjectionA1}. The zero-mode of $\phi_+^{(0)}$
transforms as a singlet under the low-energy gauge group $\SU2\times\U1$.

This example illustrates some important aspects of orbifold compactifications:
\begin{enumerate}[(i)]
\item Gauge symmetry gets reduced. \item Bulk fields furnishing
representations under the (larger) bulk gauge symmetry survive the projection
conditions only partially. That means that the same mechanism that breaks the
gauge symmetry leads to the appearance of split multiplets. 
\end{enumerate}
The  simplest setup highlighting the second point is the Kawamura model
\cite{Kawamura:1999nj,Kawamura:2000ev}.

Kawamura proposed an orbifold compactification with bulk gauge symmetry \SU5. As
we have already seen, an $\mathbbm{S}^1/\mathbbm{Z}_2$ orbifold compactification
can be equivalently described as a (field) theory on an interval whereby the
boundary conditions, imposed at the ends of the interval, can involve an
automorphism of the gauge group. In the original papers
\cite{Kawamura:1999nj,Kawamura:2000ev,Barbieri:2000vh}, orbifolds
$\mathbbm{S}^1/(\mathbbm{Z}_2\times\mathbbm{Z}_2')$ were constructed in order to
allow for different boundary conditions at the ends of the interval. In these
notes, we will not describe this mechanism in detail since the appearance of
different boundary conditions at different fixed points can be attributed to the
presence of discrete Wilson lines, as we shall see later (see subsection
\ref{sec:OrbifoldConstructionKit}). The important point is that either way one
arrives at a setup where the boundary conditions at $x_5=0$ for bulk fields in
the fundamental representation read 
\begin{subequations}
\begin{eqnarray}
 A_\mu^a(x_\mu,x_5=0)\,\mathsf{T}_a & = &
 A_\mu^a(x_\mu,x_5=0)\,P\,\mathsf{T}_a\,P^{-1}\;,\\
 \phi(x_\mu,x_5=0) & = & P\,\phi(x_\mu,x_5=0)
 \;,
\end{eqnarray}
\end{subequations}
and at $x_5=L$
\begin{subequations}
\begin{eqnarray}
 A_\mu^a(x_\mu,x_5=L)\,\mathsf{T}_a & = &
 A_\mu^a(x_\mu,x_5=L)\,P'\,\mathsf{T}_a\,P^{\prime\,-1}\;,\\
 \phi(x_\mu,x_5=L) & = & P'\,\phi(x_\mu,x_5=L)
 \;.
\end{eqnarray}
\end{subequations}
Here, we assumed that $\phi$ transforms as a $\boldsymbol{5}$-plet.

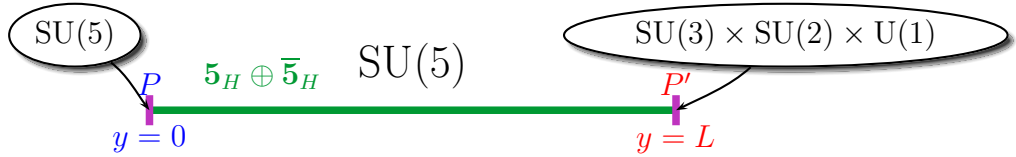
\begin{figure}[h]
\begin{center}
 \begin{pspicture}(-5,-0.5)(7,1.6)
  \psline[linewidth=1mm,linecolor=darkgreen]{-}(-3.5,0)(3.5,0)
  \psline[linewidth=1mm,linecolor=purple]{-}(-3.5,-0.2)(-3.5,0.2)
  \psline[linewidth=1mm,linecolor=purple]{-}(3.5,-0.2)(3.5,0.2)
  \pnode(-3.5,0.5){l}
  \pnode(3.5,0.5){r}
  \rput[b](-3.5,0.2){${\blue P}$}
  \rput[b](3.5,0.2){${\red P'}$}
  \rput[t](-3.5,-0.2){${\blue y=0}$}
  \rput[t](3.5,-0.2){${\red y=L}$}
  \rput[b](0,0.3){\Large $\SU5$}
  \rput[r](-3.6,1){\ovalnode[shadow=true,blur=true]{SU5}{$\SU5$}}
  \rput[l](2,1){\ovalnode[shadow=true,blur=true]{SM}{$\SU3\times\SU2\times\U1$}}
  \pnode(-3.5,0){SU51}
  \pnode(3.5,0){SM1}
  \ncarc{->}{SU5}{SU51}
  \ncarc{->}{SM}{SM1}
  \rput[c](-2,0.5){{\darkgreen $\boldsymbol{5}_H\oplus \overline{\boldsymbol{5}}_H$}}	
 \end{pspicture}
\end{center}
\caption{One-dimensional orbifold with non-trivial gauge group topography.}
\label{fig:}
\end{figure}

Denoting the eigenstates of the projections $P$ and $P'$ by $\phi_{\pm\pm}$, we
can expand these fields in terms of Fourier eigenstates (cf.\
\cite{Barbieri:2000vh}),
\begin{subequations}
\begin{eqnarray}
 \phi_{++}(x_\mu,x_5) & = &
 	\sum_{n=0}^\infty \frac{1}{\sqrt{2^{\delta_{n0}}\pi\, R}}\,
	\phi_{++}^{(2n)}(x_\mu)\,
		\cos\left(\frac{2n\,x_5}{R}\right)
 \;,\\
 \phi_{+-}(x_\mu,x_5) & = &
 	\sum_{n=0}^\infty \frac{1}{\sqrt{\pi\, R}}\,
	\phi_{+-}^{(2n+1)}(x_\mu)\,
		\cos\left(\frac{(2n+1)\,x_5}{R}\right)
 \;,\\
 \phi_{-+}(x_\mu,x_5) & = &
 	\sum_{n=0}^\infty \frac{1}{\sqrt{\pi\, R}}\,
	\phi_{-+}^{(2n+1)}(x_\mu)\,
		\sin\left(\frac{(2n+1)\,x_5}{R}\right)
 \;,\\
 \phi_{--}(x_\mu,x_5) & = &
 	\sum_{n=0}^\infty \frac{1}{\sqrt{\pi\, R}}\,
	\phi_{--}^{(2n+2)}(x_\mu)\,
		\sin\left(\frac{(2n+2)\,x_5}{R}\right)
 \;.
\end{eqnarray}
\end{subequations}
The only state that leads to a 4D zero-mode is $\phi_{++}$. That is, 4D
zero-modes are the  modes that survive all projection conditions
simultaneously.

The bulk states of the Kawamura model are the gauge bosons of \SU5 as well as
$\boldsymbol{5}$ and $\overline{\boldsymbol{5}}$ hypermultiplets. The fact that
the latter are hypermultiplets means that in 4D language each of these gives rise
to two chiral multiplets (cf.\ e.g.\ \cite{Hebecker:2001ke}). The gauge
multiplet in 5D gives rise to the usual 4D gauge fields plus a 4D chiral
superfield.

For $\boldsymbol{5}$ and $\overline{\boldsymbol{5}}$ of $\mathrm{SU}(5)$, $P$ can
be represented  by the matrix $P=\diag(1,1,1,1,1)=:\diag(+,+,+,+,+)$ and $P'$ by
$P'=\diag(+,+,-,-,-)$.\footnote{Note that with this convention, $P'$ and $-P'$
act the same way on the adjoint representation, $\ad
\SU5\sim(\boldsymbol{5}\times\overline{\boldsymbol{5}})_\mathrm{traceless}$. That is,
the action of $P'$ on the adjoint can be represented by an inner automorphism.}
The action of $P'$ is the one that breaks $\mathrm{SU}(5)$.
One easily confirms that the gauge theory on the $P'$-brane is reduced to 
$\mathrm{SU}(3)\times\mathrm{SU}(2)\times\mathrm{U}(1)$. Apart from that,
starting with a $\boldsymbol{5}$-plet in the bulk, one can repeat the analysis
of the previous example. The projection conditions are such that in the
decomposition
\begin{equation}
 \boldsymbol{5}_\mathrm{hyper}
 ~\to~
 \left[
 (\boldsymbol{3},\boldsymbol{1})_{-1/3}
 \oplus
 (\overline{\boldsymbol{3}},\boldsymbol{1})_{1/3}
 \oplus
 (\boldsymbol{1},\boldsymbol{2})_{1/2}
 \oplus
 (\boldsymbol{1},\boldsymbol{2})_{-1/2}
 \right]_\mathrm{chiral}
\end{equation}
under $\SU5$ in 5D to $G_\mathrm{SM}$ in 4D only the doublet
$(\boldsymbol{1},\boldsymbol{2})_{1/2}$ has a zero-mode. The other modes, in
particular all triplets, get heavy. The same mechanism that is used for
gauge symmetry breaking leads to doublet-triplet splitting!

Of course, in order to obtain a potentially realistic model, one also has to include
the SM matter. In the original Kawamura model \cite{Kawamura:2000ev} (and
some of its variations \cite{Altarelli:2001qj,Hall:2001pg}), SM matter states are
brane fields living at the \SU5 brane $x_5=0$. It is clear that at this point
matter has to appear in complete representations of the local, i.e.\ \SU5, gauge
symmetry. Simply stated, fields living at $x_5=0$ `do not care' for what's going
on at $x_5=L$. 

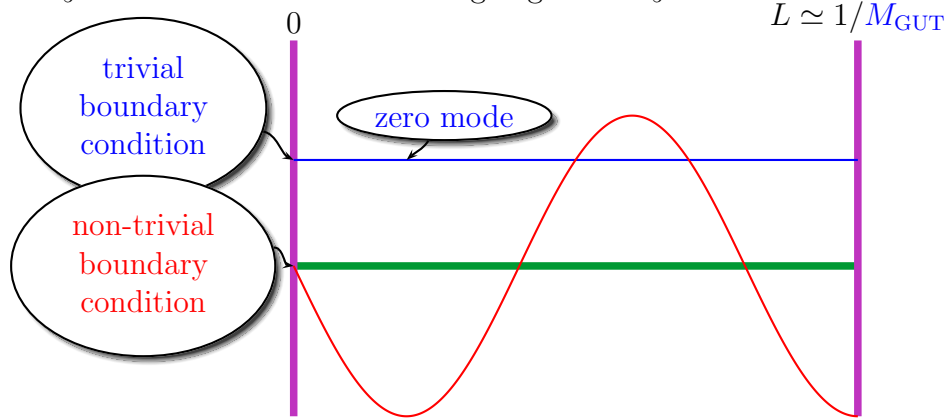
\begin{figure}[h]
\begin{center}
\begin{pspicture}(-3,-2)(8,3)
  \psset{plotpoints=1000}
  \psline[linecolor=darkgreen,linewidth=1mm]{-}(0,0)(7.5,0)
  \psline[linecolor=purple,linewidth=1mm]{-}(0,3)(0,-2)\rput[b](0,3.1){$0$}
  \psline[linecolor=purple,linewidth=1mm]{-}(7.5,3)(7.5,-2)
  \rput[b](7.5,3.1){$L\simeq1/{\blue M_\mathrm{GUT}}$}
  \psline[linecolor=blue]{-}(0,1.41)(7.5,1.41)
  \rput(2,2){\ovalnode[shadow=true,blur=true]{zm}{{\blue zero mode}}}
  \pnode(1.5,1.41){zm1}
  \ncarc{->}{zm}{zm1}
  \rput(-2,2.1){\ovalnode[shadow=true,blur=true]{tbc}{\begin{tabular}{c}
 	{\blue trivial}\\
	{\blue boundary}\\
	{\blue condition}\end{tabular}}}
  \pnode(0,1.41){tbc1}	
  \ncarc{->}{tbc}{tbc1}
%   \infixtoRPN{2*sin(-x*180/3)}
%   \typeout{dummel}
%   \typeout{\RPN}
  \psplot[linecolor=red]{0}{7.5}{2 x neg 180 mul 3 div sin mul}
  \rput(-2,-0.){\ovalnode[shadow=true,blur=true]{nbc}{\begin{tabular}{c}
 	{\red non-trivial}\\
	{\red boundary}\\
	{\red condition}\end{tabular}}}
  \pnode(0,0){nbc1}	
  \ncarc{->}{nbc}{nbc1}
\end{pspicture}
\end{center}
\caption{Gauge symmetry breaking by boundary conditions.}
\label{fig:GaugeSymmetryBreaking}
\end{figure}

What one can learn from the example is that higher-dimensional models offer an
intuitive explanation of the presence of complete and split multiplets at 
the same time.

Furthermore, it has been shown that the dimension 5 proton decay  operators, which usually lead
 to problems in 4D SUSY GUTs, are naturally absent in higher
dimensions. If the Higgs multiplets live in the bulk, the mass partners of the
triplets do not couple to SM matter, and therefore the triplet exchange diagram
does not exist \cite{Altarelli:2001qj,Hall:2001pg}. Another possibility (which,
however, ruins the intuitive explanation of SM matter in terms of complete \SU5
representations) is to introduce the Higgs on the $G_\mathrm{SM}$ brane at
$x_5=L$. Here, it is not necessary to introduce the triplet at all, and
therefore the diagram does not exist either \cite{Hebecker:2001wq}. Although
similar suppression mechanisms exist in 4D models (cf.\ \cite{Babu:1993we}), it
is intriguing that the dimension 5 proton decay problem can be solved so easily
in higher-dimensional settings.

There are further lessons to  be learned by going beyond 5D.

\subsection{Higher-dimensional orbifolds}

Let us now turn to the discussion of higher-dimensional, field-theoretic
orbifolds.
At this point, it is advantageous to introduce the usual orbifold language
\cite{Dixon:1985jw,Dixon:1986jc}. We are interested in orbifolds that emerge from
dividing a $d$-dimensional torus $\mathbbm{T}^d$ by one of its symmetries. The
torus $\mathbbm{T}^d$ can be understood as $\mathbbm{R}^d/\Gamma$, i.e.\ as the
$d$-dimensional space with points differing by lattice vectors $e_\alpha \in
\Gamma$ identified. The lattice $\Gamma$ respects a reflection symmetry, which
maps $\Gamma$ onto itself. We will often be interested in cases where the lattice
enjoys discrete rotational symmetries, i.e.\ there exists an automorphism
$\theta\in\mathrm{O}(d)$ of the lattice with $\theta^N=\mathbbm{1}$ for some
$N\in \mathbbm{N}$. The set of all such symmetries forms the point group
$\mathcal{P}$. We will be interested in a $\Z{N}$ orbifold, where any element of
the point group can be written as $\theta^k$ with $0\le k\le N-1$ ($N$ is the
smallest integer with $\theta^N=\mathbbm{1}$). One can define an orbifold as the
quotient $\mathbbm{O}=\mathbbm{T}^d/\mathcal{P}$. Equivalently one can describe
the orbifold by
\begin{equation}
 \mathbbm{O}~=~\frac{\mathbbm{R}^d}{\mathcal{S}}\;,
\end{equation}
where the space group $\mathcal{S}$ comprises of the elements of the point group and lattice
translations,
\begin{equation}
 \mathcal{S}~=~\left\{g;\;g=\left(\theta^k,n_\alpha\,e_\alpha\right)\right\}\;.
\end{equation}
Space group elements $g\in\mathcal{S}$ act on the coordinates of the compact
space as
\begin{equation}
 g\,x ~=~\theta^k\,x+n_\alpha\,e_\alpha\;.
\end{equation}
Two elements $g,h\in\mathcal{S}$ can be multiplied according to the rule
\begin{equation}\label{eq:SpaceGroupMultiplication}
 g\cdot h ~=~(\theta^k,n_\alpha\,e_\alpha)\cdot(\theta^\ell,m_\alpha\,e_\alpha)
 ~=~(\theta^{k+\ell},n_\alpha\,e_\alpha+\theta^k\,m_\alpha\,e_\alpha)
 \;.
\end{equation}

\begin{figure}[h]
\begin{center}
  \newgray{gray90}{0.9}
  \newgray{gray80}{0.8}
  \newgray{gray70}{0.7}
  \psset{unit=1.75cm}
\begin{pspicture}(-2.1,-0.6)(6.1,2.7)
\pspolygon[fillstyle=solid,linecolor=gray90,fillcolor=gray90](-2,-0.5)(6,-0.5)(6,2.5)(-2,2.5)(-2,-0.5)
	\psdot*[dotstyle=*,dotsize=2mm,linecolor=blue](-0.5,0)
	\psdot*[dotstyle=*,dotsize=2mm,linecolor=blue](1.5,0)
	\psdot*[dotstyle=*,dotsize=2mm,linecolor=blue](3.5,0)
	\psdot*[dotstyle=*,dotsize=2mm,linecolor=blue](5.5,0)
	\psdot*[dotstyle=*,dotsize=2mm,linecolor=blue](-1,1)
	\psdot*[dotstyle=*,dotsize=2mm,linecolor=blue](1,1)
	\psdot*[dotstyle=*,dotsize=2mm,linecolor=blue](3,1)
	\psdot*[dotstyle=*,dotsize=2mm,linecolor=blue](5,1)
	\psdot*[dotstyle=*,dotsize=2mm,linecolor=blue](-1.5,2)
	\psdot*[dotstyle=*,dotsize=2mm,linecolor=blue](0.5,2)
	\psdot*[dotstyle=*,dotsize=2mm,linecolor=blue](2.5,2)
	\psdot*[dotstyle=*,dotsize=2mm,linecolor=blue](4.5,2)
\pspolygon[fillstyle=solid,linecolor=gray80,fillcolor=gray80](1,1)(3,1)(2.5,2)(0.5,2)(1,1)
\psline[linewidth=1mm,linecolor=blue]{->}(1,1)(3,1)
\psline[linewidth=1mm,linecolor=blue]{->}(1,1)(0.5,2)
\rput[tl](3.1,0.9){{\blue $e_1$}}
\rput[br](0.4,2.1){{\blue $e_2$}}
\psline[linewidth=1mm,linecolor=blue]{->}(1,1)(3,1)
\psline[linewidth=1mm,linecolor=blue]{->}(1,1)(0.5,2)
	\psdot*[dotstyle=o,dotsize=3mm,linecolor=red](1,1)
	\psdot*[dotstyle=o,dotsize=3mm,linecolor=red](2,1)
	\psdot*[dotstyle=o,dotsize=3mm,linecolor=red](1.75,1.5)
	\psdot*[dotstyle=o,dotsize=3mm,linecolor=red](0.75,1.5)
	\psdot*[dotstyle=*,dotsize=2mm,linecolor=red](1,1)
	\psdot*[dotstyle=*,dotsize=2mm,linecolor=red](2,1)
	\psdot*[dotstyle=*,dotsize=2mm,linecolor=red](1.75,1.5)
	\psdot*[dotstyle=*,dotsize=2mm,linecolor=red](0.75,1.5)
\pspolygon[fillstyle=solid,linecolor=black,fillcolor=gray70,linestyle=dashed](1,1)(2,1)(1.5,2)(0.5,2)(1,1)
\psline[linewidth=1mm,linecolor=blue]{->}(1,1)(3,1)
\psline[linewidth=1mm,linecolor=blue]{->}(1,1)(0.5,2)
	\psdot*[dotstyle=o,dotsize=3mm,linecolor=red](1,1)
	\psdot*[dotstyle=o,dotsize=3mm,linecolor=red](2,1)
	\psdot*[dotstyle=o,dotsize=3mm,linecolor=red](1.75,1.5)
	\psdot*[dotstyle=o,dotsize=3mm,linecolor=red](0.75,1.5)
	\psdot*[dotstyle=*,dotsize=2mm,linecolor=red](1,1)
	\psdot*[dotstyle=*,dotsize=2mm,linecolor=red](2,1)
	\psdot*[dotstyle=*,dotsize=2mm,linecolor=red](1.75,1.5)
	\psdot*[dotstyle=*,dotsize=2mm,linecolor=red](0.75,1.5)
	\pnode(1.25,1.5){fdo1}
	\ncarc{->}{fdo}{fdo1}
\psarc[linecolor=purple,linewidth=1mm]{->}(1,1){0.7}{60}{240}
\end{pspicture}\psset{unit=1cm}
\end{center}  
\caption{$\mathbbm{T}^2/\mathbbm{Z}_2$ orbifold.}
\label{fig:T2overZ2}
\end{figure}
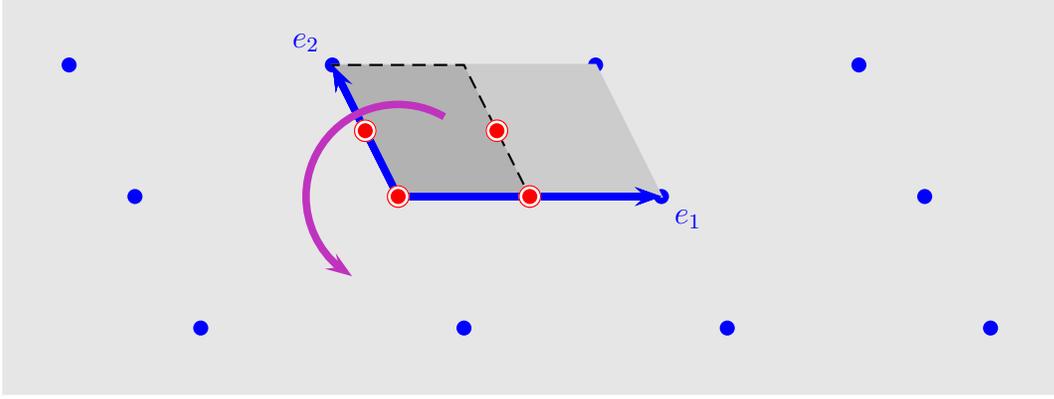

Let us first consider the orbifold $\mathbbm{T}^2/\mathbbm{Z}_2$, which emerges
from dividing the torus $\mathbbm{T}^2$ by a point reflection symmetry. The
torus is defined by  2 (linearly independent) lattice vectors $e_1$ and $e_2$
spanning the fundamental domain (in figure~\ref{fig:T2overZ2} the fundamental
domain is given by the two shaded regions). The $\Z2$ then acts as a reflection
(or, equivalently, a  $180^\circ$ rotation represented by the purple arc) about
an arbitrary lattice node which one could call the `origin'. Certain points are
mapped under the orbifold action onto themselves (up to lattice translations);
these are the orbifold fixed points (red bullets). By identifying points in the
fundamental domain of the torus, that are related by the \Z2 orbifold action,
one arrives at the fundamental domain of the orbifold. In
figure~\ref{fig:T2overZ2} this is the darker shaded region. By folding the
fundamental domain along the line connecting the upper two fixed points and
gluing the adjacent edges together, one arrives at a ravioli- (cf.\
\cite{Quevedo:1996sv}) or pillow- (cf.\ \cite{Hebecker:2003jt}) like object
which is smooth everywhere except for the corners, i.e.\ the orbifold fixed
points.

\begin{figure}[h]
\begin{center}
\CenterObject{\psset{unit=1.2cm}
\begin{pspicture}(-2,-0.8)(6.5,4.5)
\pscustom[slopesteps=200,fillstyle=ccslope,slopebegin=white,slopeend=gray,slopecenter=0.7 0.7]{
	\pscurve[liftpen=1](0,0)(2,0.2)(4,0)
	\pscurve[liftpen=1](4,0)(3.7,2)(4,4)
	\pscurve[liftpen=1](4,4)(2,3.8)(0,4)
	\pscurve[liftpen=1](0,4)(0.3,2)(0,0)
	}
	\pscurve[linewidth=0.5mm](0.8,0.28)(2,0.4)(3.2,0.28)
	\pscurve[linewidth=0.5mm](1.2,0.5)(2,0.56)(2.8,0.5)
	\pscurve[linewidth=0.5mm](0.8,3.72)(2,3.6)(3.2,3.72)
	\pscurve[linewidth=0.5mm](1.2,3.5)(2,3.44)(2.8,3.5)
	\pscurve[linewidth=0.5mm](0.36,0.8)(0.48,2)(0.36,3.2)
	\pscurve[linewidth=0.5mm](0.6,1.2)(0.68,2)(0.6,2.8)
	\pscurve[linewidth=0.5mm](3.64,0.8)(3.52,2)(3.64,3.2)
	\pscurve[linewidth=0.5mm](3.4,1.2)(3.32,2)(3.4,2.8)
	\psdot*[dotstyle=o,dotsize=3mm,linecolor=gray](0,0)
 	\psdot*[dotstyle=square*,dotsize=2mm,linecolor=gray](0,0)
	\psdot*[dotstyle=o,dotsize=3mm,linecolor=gray](4,0)
	\psdot*[dotstyle=square*,dotsize=2mm,linecolor=gray](4,0)
	\psdot*[dotstyle=o,dotsize=3mm,linecolor=gray](4,4)
	\psdot*[dotstyle=square*,dotsize=2mm,linecolor=gray](4,4)
	\psdot*[dotstyle=o,dotsize=3mm,linecolor=gray](0,4)
	\psdot*[dotstyle=square*,dotsize=2mm,linecolor=gray](0,4)
	\rput[c](2,2){\begin{tabular}{c}
	{\Huge $\SO{10}$}\end{tabular}}
	\pnode(2.5,1.2){bulk}
	\psdot*[dotstyle=o,dotsize=3mm,linecolor=blue](0,0)
 	\psdot*[dotstyle=square*,dotsize=2mm,linecolor=blue](0,0)
	\pnode(0,0){lu}
\rput[br](0,0.2){\ovalnode[shadow=true,blur=true]{P10}{{\blue $P=\mathbbm{1}_{10}$}}}
\rput[rt](0.5,-0.2){\rnode[c]{SO10}{\large{\blue$\SO{10}$}}}
\ncarc{->}{P10}{lu}
	\psdot*[dotstyle=o,dotsize=3mm,linecolor=darkgreen](4,0)
	\psdot*[dotstyle=square*,dotsize=2mm,linecolor=darkgreen](4,0)
	\pnode(4,0){ru}
\rput[b](4.9,0.2){%
\ovalnode[shadow=true,blur=true]{PGG}{{\darkgreen $P_\mathrm{GG}=\diag(\sigma_2,\sigma_2,\sigma_2,\sigma_2,\sigma_2)$}}}
	\rput[lt](4,-0.2){$\rnode[c]{Grb}{{\darkgreen \SU5\times\U1}}$}
\ncarc{->}{PGG}{ru}
	\psdot*[dotstyle=o,dotsize=3mm,linecolor=red](4,4)
	\psdot*[dotstyle=square*,dotsize=2mm,linecolor=red](4,4)
	\pnode(4,4){ro}
	\rput[t](4,3.8){%
\ovalnode[shadow=true,blur=true]{PPS}{{\red $P_\mathrm{PS}=\diag(-\mathbbm{1}_2,-\mathbbm{1}_2,-\mathbbm{1}_2,\mathbbm{1}_2,\mathbbm{1}_2)$}}}
\rput[b](4,4.2){${\red G_\mathrm{PS}=\SU4\times\SU2_\mathrm{L}\times\SU2_\mathrm{R}}$}
\ncarc{->}{PPS}{ro}
	\psdot*[dotstyle=o,dotsize=3mm,linecolor=purple](0,4)
	\psdot*[dotstyle=square*,dotsize=2mm,linecolor=purple](0,4)
	\pnode(0,4){lo}	
	\rput[rb](-0.2,4.2){${\purple G_\mathrm{fl}}$}
	\rput[tr](0,3.9){%
\ovalnode[shadow=true,blur=true]{Pfl}{${\purple P_\mathrm{fl}}=
{\red P_\mathrm{PS}}\cdot{\darkgreen P_\mathrm{GG}}$}}
\ncarc{->}{Pfl}{lo}
\end{pspicture}\psset{unit=1.2cm}}
\end{center}
\label{fig:ABCmodel}
\caption{Asaka-Buchm\"uller-Covi model.}
\end{figure}
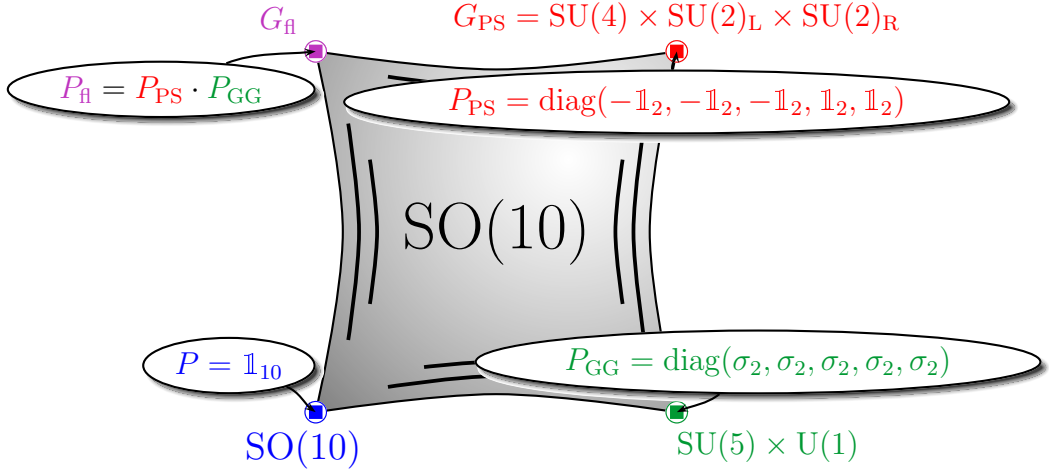

As a specific example of a 6D orbifold, let us discuss the
Asaka-Buchm\"uller-Covi model \cite{Asaka:2001eh,Asaka:2002nd}. Here, the gauge
embedding (`orbifold parities') is chosen such that \SO{10} is broken to
$G_\mathrm{GG}=\SU5\times\U1$, the Pati-Salam symmetry $G_\mathrm{PS}$ and the
so-called flipped \SU5 $G_\mathrm{fl}$ at three of the four fixed points while
the gauge embedding is trivial at the fourth fixed point (cf.\ figure
\ref{fig:ABCmodel}). It is a remarkable (group-theoretic) fact that the
intersection of $G_\mathrm{GG}$ and $G_\mathrm{PS}$ in \SO{10} yields the
standard model, $G_\mathrm{SM}$, up to a \U1 factor. In order to accommodate
the observed three generations, the authors of \cite{Asaka:2001eh,Asaka:2002nd}
placed one generation at each of  the $G_\mathrm{GG}$, $G_\mathrm{PS}$ and
$G_\mathrm{fl}$ fixed points. Clearly, those localized fields have to furnish
complete representations of the `local' gauge groups realized at the fixed
points. The Higgs arises from the bulk $\boldsymbol{10}$-plets whereby only the
electroweak doublets survive all projection conditions and  have a zero-mode.

There are certain lessons that can be learned from 2-dimensional orbifolds:
\begin{enumerate}
 \item these constructions can exhibit a `non-trivial gauge group topography',
 i.e.\ a bulk group gets broken to different `local groups' at different fixed
 points;
 \item the low-energy gauge group is the intersection of the local groups in the
 bulk group;
 \item localized matter comes in complete representations of the local gauge
 groups;
 \item bulk fields get split, i.e. are partially projected out. 
\end{enumerate}

Nevertheless, it is also clear that these models leave many questions
unanswered. One would, for instance, like to obtain a deeper  understanding  of
the field content, i.e.\ the bulk fields and the states localized at the fixed
points. Further, it would be useful  to have a theory behind the couplings
among various fields. It is also clear that calculations based on
higher-dimensional field theories (which are known to be non-renormalizable)
cannot be fully trusted and require a UV completion.  Gravity should also be
incorporated. Last but not least, the orbifold GUTs in the literature provide a
nice explanation of the fact that the Higgses appear  as split multiplets,
however the fact that matter generations fit into $\boldsymbol{16}$-plets of
\SO{10} has no convincing explanation. In what follows, we will explain how 
string-derived orbifolds provide a framework to address all these questions.

\section{Orbifold compactifications of the heterotic string}
\label{sec:StringOrbifolds}

Orbifold compactifications of the heterotic string have a long history
\cite{Dixon:1985jw,Dixon:1986jc,Ibanez:1986tp,Ibanez:1987sn,Casas:1987us,Casas:1988hb,Font:1988tp,Font:1988nc,Font:1988mm,Font:1989aj}
and remain the most phenomenologically  attractive  string framework. 
As is well-known, there are 5 string theories. Since we aim at obtaining
$\boldsymbol{16}$-plets of \SO{10} (at the perturbative level) as well as gauge unification, the heterotic
string \cite{Gross:1984dd,Gross:1985fr} is singled out. 
A  natural setup accommodating  these features is   the $\E8\times\E8$ heterotic
string, which will be our  focus in what follows.
Introductory lectures on $\E8\times\E8$ heterotic orbifolds can be found in \cite{Ibanez:1987dw}.

The first string-derived orbifold GUTs were presented in
\cite{Kobayashi:2004ud,Forste:2004ie,Kobayashi:2004ya,Buchmuller:2004hv}. While in
\cite{Forste:2004ie} the stringy calculation of the spectrum and the connection
between Wilson lines and non-trivial gauge group topographies are explained in
great detail, Ref.~\cite{Kobayashi:2004ud,Kobayashi:2004ya}  provides an orbifold
GUT interpretation of  heterotic orbifolds  and derives    models with
Pati-Salam gauge symmetry and three chiral generations of matter. Further,
Ref.~\cite{Buchmuller:2004hv} presents an MSSM--like model with 3 localized  $\boldsymbol{16}$-plets
and studies its orbifold GUT limits in various (5 to 9)  dimensions.  
These  models are based on a \Z6-II orbifold, which we describe in the next subsection.

\subsection{Orbifold basics (using the $\boldsymbol{\Z6}$-II orbifold)}

The compactification lattice can be chosen as (see figure
\ref{fig:LambdaG2xSU3xSO4})
\begin{equation}
\Lambda_{\G2 \times \SU3 \times \SO4}
~:=~\text{root lattice of Lie algebra of }\:\G2 \times \SU3 \times \SO4\;.
\end{equation}

\begin{figure}[h]
\centerline{
\CenterObject{\psset{unit=1cm}
\begin{pspicture}(-1.5,-0.1)(6,2.1)
   	\pspolygon[fillstyle=solid,fillcolor=gray90,linecolor=gray80]
		(0,0)(2.309,0)(5.773,2)(3.464,2)
   	\psline[linewidth=0.8mm,linecolor=red]{*->}(0,0)(3.464,2)
   	\psline[linewidth=0.8mm,linecolor=red]{*->}(0,0)(2.309,0)
	\rput[t](2.2,-0.1){\small {\red $\re z_1$}}
\psarc[linecolor=purple,linewidth=1mm]{<-}(0,0){1}{315}{15}
	\psdot*[dotstyle=o,dotsize=3mm](0,0)
	\psdot*[dotstyle=*,dotsize=2mm](0,0)
\psdot*[dotstyle=o,dotsize=3mm](-1.4,2.1)
\psdot*[dotstyle=*,dotsize=2mm](-1.4,2.1)
\rput[l](-1.1,2.1){=$\mathbbm{Z}_6$ fixed point}
\end{pspicture}
\psset{unit=1cm}}
$\times$
\CenterObject{\psset{unit=1.2cm}
\begin{pspicture}(-1.2,-0.1)(2,1.8)
	\pspolygon[fillstyle=solid,fillcolor=gray90,linecolor=gray80]
		(0,0)(2,0)(1,1.73205)(-1,1.73205)(0,0)
	\psline[linewidth=0.8mm,linecolor=darkgreen]{*->}(0,0)(2,0)
	\psline[linewidth=0.8mm,linecolor=darkgreen]{*->}(0,0)(-1,1.73205)
	\rput[t](1.8,-0.1){\small {\darkgreen $\re z_2$}}
\psarc[linecolor=purple,linewidth=1mm]{<-}(0,0){1}{300}{60}
	\psdot*[dotstyle=o,dotsize=3mm](0,0)
 	\psdot*[dotstyle=*,dotsize=2mm](0,0)
	\psdot*[dotstyle=o,dotsize=3mm](1,0.5)
	\psdot*[dotstyle=*,dotsize=2mm](1,0.5)
	\psdot*[dotstyle=o,dotsize=3mm](0.0669873, 1.11603)
	\psdot*[dotstyle=*,dotsize=2mm](0.0669873, 1.11603)
\end{pspicture}\psset{unit=1cm}}
$\times$
\CenterObject{\psset{unit=1.2cm}
\begin{pspicture}(-0.6,-0.1)(2.1,2.1)
	\pspolygon[fillstyle=solid,fillcolor=gray90,linecolor=gray80]
		(0,0)(2,0)(2,2)(0,2)(0,0)
	\psline[linewidth=0.8mm,linecolor=blue]{*->}(0,0)(2,0)
	\rput[t](1.8,-0.1){\small {\blue $\re z_3$}}
	\psline[linewidth=0.8mm,linecolor=blue]{*->}(0,0)(0,2)
\psarc[linecolor=purple,linewidth=1mm]{->}(0,0){1}{45}{225}
	\psdot*[dotstyle=o,dotsize=3mm](0,0)
 	\psdot*[dotstyle=*,dotsize=2mm](0,0)
	\psdot*[dotstyle=o,dotsize=3mm](1,0)
	\psdot*[dotstyle=*,dotsize=2mm](1,0)
	\psdot*[dotstyle=o,dotsize=3mm](1,1)
	\psdot*[dotstyle=*,dotsize=2mm](1,1)
	\psdot*[dotstyle=o,dotsize=3mm](0,1)
	\psdot*[dotstyle=*,dotsize=2mm](0,1)
\end{pspicture}\psset{unit=1cm}}
}
\caption{$\Lambda_{\G2\times\SU3\times\SO4}$.}
\label{fig:LambdaG2xSU3xSO4}
\end{figure}

The \Z6 action is a $-60^\circ$ rotation in the \G2 plane, a $-120^\circ$ rotation
in the \SU3 plane and a $180^\circ$ rotation (or reflection) in the \SO4 plane.
Parametrizing the three 2-tori by complex coordinates $z_i$ ($1\le i\le3$), the
orbifold action on the compact six dimensions reads
\begin{equation}
 z_i~\rightarrow~e^{2\pi \I\, v_6^i}~ z_i
\quad\text{with}\qquad
v_6~=~\frac{1}{6}(-1,-2,3)\;.
\end{equation}
This orbifold twist  is  accompanied by the corresponding  action on the gauge degrees of
freedom. Using the string notation, one has (cf.\
\cite{Forste:2004ie,Buchmuller:2006ik})
\begin{equation}
 (\theta,0)\quad : \quad X^I~\to~X^I+\pi\, V_6^I\;,
\end{equation}
where $(\theta,0)\in\mathcal{S}$, $6\,V_6\in\Lambda_{\E8\times\E8}$ for the
action to be \Z6. Here, $X^I$ ($1\le I \le 16$) denote the left-moving string
coordinates which are compactified on the $\E8\times\E8$ torus
$\Lambda_{\E8\times\E8}$. Moreover, the torus translations can be associated with the
so-called Wilson lines \cite{Ibanez:1986tp}, e.g.\
\begin{equation}\label{eq:WilsonTranslation}
 (\mathbbm{1},e_5)\quad : \quad z_3~ \to~z_3+1\quad\leftrightarrow\quad
 X^I~\to~ X^I+\pi\, W_2\;,
\end{equation}
where $2\,W_2\in\Lambda_{\E8\times\E8}$. Wilson lines are subject to certain
constraints; the \Z6-II orbifold discussed here allows for two 
Wilson lines of order 2 associated with  the two independent translations in the
\SO4 torus and one Wilson line of order 3  in the \SU3 torus ($3\,W_3\in\Lambda_{\E8\times\E8}$).

\subsection{Orbifold construction kit}
\label{sec:OrbifoldConstructionKit}

\begin{figure}[h]
\begin{pspicture}(-4,-2.8)(8,3.2)
\rput[rt](0,1){\CornerA{0}{lt}{rb}{%
	\rnode[c]{V1}{$V_\mathrm{bl}$}}{%
	\rnode[c]{G1}{$G_\mathrm{bl}$}}{0.2}}
\psset{linecolor=red}
\rput[rb]{-90}(-3,1){\CornerA{90}{bl}{rt}{%
	${\red V_\mathrm{tl}}$}{%
	${\red G_\mathrm{tl}}$}{-0.1}}
\psset{linecolor=blue}
\rput[lb]{-180}(4,4.1){\CornerA{180}{tl}{rb}{%
	${\blue V_\mathrm{tr}}$}{%
	${\blue G_\mathrm{tr}}$}{-0.2}}
\psset{linecolor=darkgreen}
\rput[lb]{-270}(4,-2.8){\CornerA{270}{lb}{rt}{%
	${\darkgreen V_\mathrm{br}}$}{%
	${\darkgreen G_\mathrm{br}}$}{0.2}}
\psset{linecolor=black}

\rput[c]{0}(8,1){\QuadraticPillow{%
$V_\mathrm{bl}=V\qquad\qquad$}{$G_\mathrm{bl}$}{%
${\darkgreen V_\mathrm{br}}=V+W'$}{${\darkgreen G_\mathrm{br}}$}{%
${\red V_\mathrm{tl}}=V+W$}{${\red G_\mathrm{tl}}$}{%
${\blue V_\mathrm{tr}}=V+W+W'$}{${\blue G_\mathrm{tr}}$}{%
$\E8\times\E8$}
}
\end{pspicture}
\caption{A 2D sketch of the orbifold construction kit. The gauge group after compactification is 
${\purple G_\mathrm{low-energy}}=G_\mathrm{bl}\cap 
{\red G_\mathrm{tl}}\cap {\darkgreen G_\mathrm{br}}\cap {\blue G_\mathrm{tr}}$.
}
\label{fig:OrbifoldConstructionKit}
\end{figure}
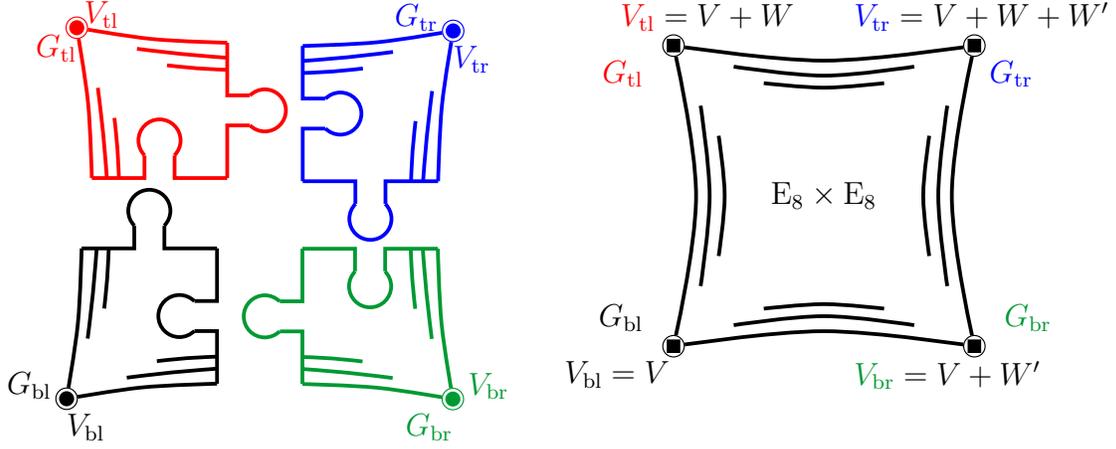

The presence of Wilson lines leads to the  picture where, as in the 6D orbifolds,
the bulk gauge group gets broken to different subgroups at different fixed points. 
Each symmetry breakdown
can be attributed to the \emph{local gauge  shift}  of the form
\begin{equation}
 V_\mathrm{local}~=~V_6+\text{Wilson lines}
 \qquad\leftrightarrow\qquad
 G_\mathrm{local}
 \;.
\end{equation}  
Here, the local gauge group $ G_\mathrm{local}$   corresponds to the  gauge
bosons which fulfill $p\cdot V_\mathrm{local}\in\mathbbm{Z}$ (see 
\cite{Buchmuller:2006ik} for details). Gauge interactions surviving
compactification are mediated by the gauge bosons that satisfy  all projection
conditions  simultaneously. In other words, they correspond to the  intersection
of all local gauge groups in the bulk group. This leads to an alternative way of
constructing orbifolds with Wilson lines: start with a local shift, i.e.\ with a
`corner' of the fundamental domain, and glue it to the other corners.  The
result is a `pillow'  with non-trivial gauge group topography (figure
\ref{fig:OrbifoldConstructionKit}). It is important that, due to stringy
constraints,  this procedure does not allow for arbitrary gauge groups at
different corners. In particular, modular invariance \cite{Vafa:1986wx}, which
guarantees anomaly freedom,  restricts possible choices of the gauge shifts and
Wilson lines. 

\subsection{Space group versus fixed points}

Space group elements enjoy the group multiplication  law \eqref{eq:SpaceGroupMultiplication}.
Hence, they  come in  equivalence classes. Two space group
elements $g$ and $g'$ are in the same class  if there is an
element $h\in \mathcal{S}$ such that
\begin{equation}
 g~=~h\,g'\,h^{-1}\;.
\end{equation}
Each conjugacy class corresponds to one fixed point in the fundamental domain. In
non-prime orbifolds such as the \Z6-II orbifold, the situation is more complicated
since there are also fixed planes  (see
\cite{Buchmuller:2006ik} for details).

\subsection{Spectrum}

\begin{figure}[h]
\centerline{
\subfigure[Twisted and untwisted strings.\label{fig:UnTwistedStrings}]{\begin{minipage}{5cm}
\CenterObject{\begin{pspicture}(-1,0)(4,-4.2)
\rput[lt](-0.4,-0.25){\CenterObject{\psset{unit=2cm}
\begin{pspicture}(0,0)(2,2)
\pscustom[slopesteps=200,fillstyle=ccslope,slopebegin=white,slopeend=gray,slopecenter=0.7 0.7]{
	\pscurve[liftpen=1](0,0)(1,0.1)(2,0)
	\pscurve[liftpen=1](2,0)(1.85,1)(2,2)
	\pscurve[liftpen=1](2,2)(1,1.9)(0,2)
	\pscurve[liftpen=1](0,2)(0.15,1)(0,0)
	}
	\pscurve(0.4,0.14)(1,0.2)(1.6,0.14)
	\pscurve(0.6,0.25)(1,0.28)(1.4,0.25)
	\pscurve(0.4,1.86)(1,1.8)(1.6,1.86)
	\pscurve(0.6,1.75)(1,1.72)(1.4,1.75)
	\pscurve(0.18,0.4)(0.24,1)(0.18,1.6)
	\pscurve(0.3,0.6)(0.34,1)(0.3,1.4)
	\pscurve(1.82,0.4)(1.76,1)(1.82,1.6)
	\pscurve(1.7,0.6)(1.66,1)(1.7,1.4)
	\psdot*[dotstyle=o,dotsize=3mm](0,0)
 	\psdot*[dotstyle=*,dotsize=2mm](0,0)
	\psdot*[dotstyle=o,dotsize=3mm](2,0)
	\psdot*[dotstyle=*,dotsize=2mm](2,0)
	\psdot*[dotstyle=o,dotsize=3mm](2,2)
	\psdot*[dotstyle=*,dotsize=2mm](2,2)
	\psdot*[dotstyle=o,dotsize=3mm](0,2)
	\psdot*[dotstyle=*,dotsize=2mm](0,2)
\end{pspicture}}}
\rput[lt](-8.9,-3.58){\psset{unit=1.25cm}
\begin{pspicture}(-2.1,-2.1)(2.1,2.1)
\psset{linecolor=red,linewidth=0.5mm}
\psset{linestyle=solid}
\pscurve(4.69,1.98)(4.81,2.01)(4.86,1.91)(5.05,1.93)(5.00,1.82)(5.05,1.74)(4.95,1.66)(5.00,1.59)
\psset{linestyle=solid}
\pscurve(5.06,4.72)(5.08,4.60)(4.98,4.56)(5.01,4.37)(4.89,4.41)(4.81,4.37)(4.73,4.46)(4.67,4.42)
\psset{linestyle=solid,linecolor=darkgreen}
\psccurve(5.45,3.98)(5.41,3.83)(5.27,3.72)(5.38,3.59)(5.44,3.39)(5.62,3.59)(5.80,3.51)(5.84,3.71)(5.89,3.88)(5.69,3.91)(5.52,4.05)
\psset{linecolor=darkgreen}
\psccurve(6.87,2.90)(6.77,2.93)(6.70,3.02)(6.61,2.94)(6.48,2.90)(6.61,2.79)(6.56,2.66)(6.69,2.64)(6.80,2.61)(6.82,2.74)(6.92,2.86)
\psccurve(6.10,2.82)(6.07,2.92)(5.97,2.99)(6.05,3.08)(6.09,3.21)(6.20,3.08)(6.33,3.13)(6.35,3.00)(6.39,2.89)(6.25,2.87)(6.14,2.77)
\end{pspicture}\psset{unit=1cm}}
\end{pspicture}}\end{minipage}}
\quad
\subfigure[Dictionary.\label{fig:Dictionary}]{\begin{minipage}{6cm}\begin{tabular}{p{2cm}|p{4cm}}
$k$ & interpretation\\
\hline
{\darkgreen $k=0$} & bulk fields\\[0.2cm]
${\red k=1,5}$ & fields live on points in 6D compact space\\[0.2cm]
${\red k=2,3,4}$ & fields live on 2-dimensional planes in 6D compact space
\end{tabular}\end{minipage}
}}
\caption{Bulk and brane fields from untwisted and twisted strings.}
\label{fig:UntwistedAndTwistedStrings}
\end{figure}
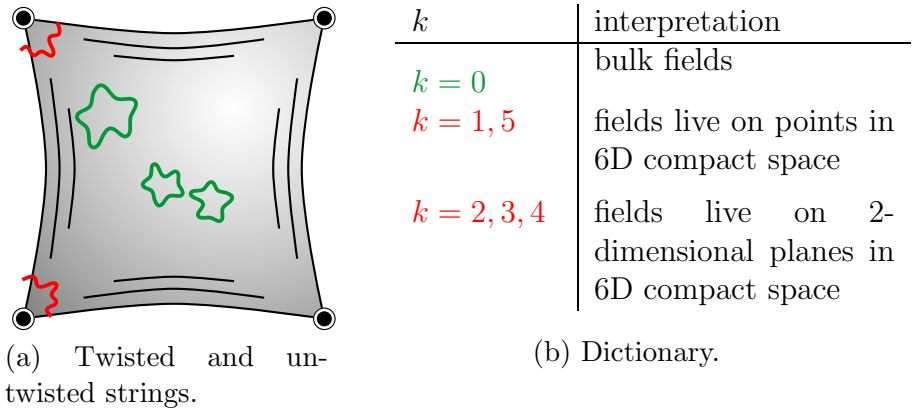

We are interested in the zero-modes on orbifolds. Each state can be associated
with a conjugacy class, or a space group element $(\theta^k,n_\alpha\,e_\alpha)$,
which will be referred to as the `constructing element'. The (zero-mode)
spectrum decomposes into an untwisted ($k=0$) and various twisted ($k\ne 0$)
sectors. All massless states 
correspond to closed strings, while  twisted states correspond to strings that
are  closed  only upon  orbifolding (figure
\ref{fig:UnTwistedStrings}). One associates
\[
 \text{massless state}
 ~\leftrightarrow~
 \text{constructing element }(\theta^k,n_\alpha\,e_\alpha)
 ~\leftrightarrow~
 \left\{\begin{array}{lcl}
 \text{fixed point} & :& k~=~1,5\;,\\
 \text{fixed plane} & : & k~=~2,3,4\;,\\
 \text{bulk} & : & k~=~0\;.
 \end{array}\right.
\]
This amounts  to a dictionary between constructing elements and  localization
properties of the corresponding state in the field-theoretic language (figure
\ref{fig:Dictionary}).  Given the geometry and gauge embedding (i.e.\ the gauge
shift and Wilson lines), one can calculate the spectrum following a
straightforward procedure  (cf.~\cite{Forste:2004ie}). For the \Z6-II
calculation, see \cite{Kobayashi:2004ya,Buchmuller:2006ik}.

\subsection{Selection rules}

Given the field content of the orbifold, the next step is to study 
interactions of the theory. Couplings on orbifolds are governed by certain
selection rules \cite{Hamidi:1986vh,Dixon:1986qv}. From the effective field
theory perspective, these rules are
\begin{itemize}
 \item gauge invariance,
 \item (non-Abelian) discrete symmetries, and
 \item discrete $R$-symmetries.
\end{itemize}
The selection rules for the \Z6-II orbifold are summarized in
\cite{Buchmuller:2006ik} (a careful discussion of the space-group rule
is presented in \cite{Lebedev:2007hv}).

\subsection{Summary of orbifold basics}

The main lesson  is
that, once the geometry and gauge embedding are fixed, the theory is completely
determined. In particular, unlike in the field-theoretic constructions, the
spectrum is calculable and one cannot `invent' representations and/or couplings.

\section{Local grand unification}
\label{sec:LGU}

Having discussed how to compactify the heterotic string on an orbifold, let us
now turn to the applications --  construction of potentially realistic string
models.

\subsection{First string-derived orbifold GUT}

 The first constructions where the intuition gained in
orbifold GUTs was used to choose appropriate geometry and gauge embeddings were
the Pati-Salam models by Kobayashi, Raby and Zhang (KRZ)
\cite{Kobayashi:2004ud,Kobayashi:2004ya}. KRZ constructed a model with the
following interesting features:
\begin{itemize}
 \item gauge group upon compactification is $G_\mathrm{PS}$;
 \item spectrum is 3  generations plus vector-like matter under $G_\mathrm{PS}$;
 \item Pati-Salam Higgses, necessary to break $G_\mathrm{PS}$ to
  $G_\mathrm{SM}$, are present in the spectrum;
 \item two generations live at fixed points with an \SO{10} GUT symmetry;
 \item many exotics can be given  masses by assigning vacuum expectation values
 (vevs) to Pati-Salam singlets.
\end{itemize}
In addition to these positive features, there are a number of shortcomings.
Perhaps, the main issue is the breaking of the Pati-Salam gauge symmetry to that
of the Standard Model. In string theory on cannot write down an arbitrary Higgs
potential -- it must be derived. In particular, it must be consistent with
various selection rules (as well as supersymmetry).  

Some of these problems can be avoided by breaking the 10D gauge symmetry 
directly to that of the SM (times extra factors) upon compactification.
This strategy allows one to construct  many models with the exact MSSM spectrum 
while keeping nice features of GUTs.

\subsection{Local grand unification}

The successful aspects of the KRZ model have triggered further
investigations. In particular, it has been pointed out that
$\boldsymbol{16}$-plets localized at the fixed points with \SO{10} symmetry can be a
very good starting point for the construction of realistic models \cite{Buchmuller:2004hv}. 
These states are not split by the orbifold projection and all survive in the 4D spectrum,
thereby providing complete matter generations of the SM. It is important that they
form a GUT representation without having GUT symmetry in 4D.

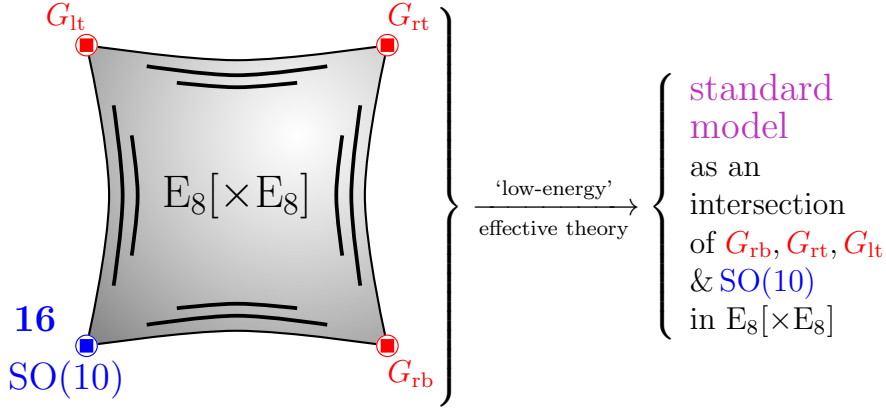
\begin{figure}[h]
\begin{center}
$\left.\text{
\CenterObject{
\begin{pspicture}(-0.1,-0.8)(4.5,4.5)
\pscustom[slopesteps=200,fillstyle=ccslope,slopebegin=white,slopeend=gray,slopecenter=0.7 0.7]{
	\pscurve[liftpen=1](0,0)(2,0.2)(4,0)
	\pscurve[liftpen=1](4,0)(3.7,2)(4,4)
	\pscurve[liftpen=1](4,4)(2,3.8)(0,4)
	\pscurve[liftpen=1](0,4)(0.3,2)(0,0)
	}
	\pscurve[linewidth=0.5mm](0.8,0.28)(2,0.4)(3.2,0.28)
	\pscurve[linewidth=0.5mm](1.2,0.5)(2,0.56)(2.8,0.5)
	\pscurve[linewidth=0.5mm](0.8,3.72)(2,3.6)(3.2,3.72)
	\pscurve[linewidth=0.5mm](1.2,3.5)(2,3.44)(2.8,3.5)
	\pscurve[linewidth=0.5mm](0.36,0.8)(0.48,2)(0.36,3.2)
	\pscurve[linewidth=0.5mm](0.6,1.2)(0.68,2)(0.6,2.8)
	\pscurve[linewidth=0.5mm](3.64,0.8)(3.52,2)(3.64,3.2)
	\pscurve[linewidth=0.5mm](3.4,1.2)(3.32,2)(3.4,2.8)
	\psdot*[dotstyle=o,dotsize=3mm,linecolor=blue](0,0)
 	\psdot*[dotstyle=square*,dotsize=2mm,linecolor=blue](0,0)
	\rput[rt](0.5,-0.2){\rnode[c]{SO10}{\large{\blue $\SO{10}$}}}
	\rput[rb](-0.4,0.2){\large{\blue $\boldsymbol{16}$}}
	\psdot*[dotstyle=o,dotsize=3mm,linecolor=red](4,0)
	\psdot*[dotstyle=square*,dotsize=2mm,linecolor=red](4,0)
	\rput[lt](4,-0.2){$\rnode[c]{Grb}{{\red G_\mathrm{rb}}}$}
	\psdot*[dotstyle=o,dotsize=3mm,linecolor=red](4,4)
	\psdot*[dotstyle=square*,dotsize=2mm,linecolor=red](4,4)
	\rput[rb](0,4.2){${\red G_\mathrm{lt}}$}
	\psdot*[dotstyle=o,dotsize=3mm,linecolor=red](0,4)
	\psdot*[dotstyle=square*,dotsize=2mm,linecolor=red](0,4)
	\rput[lb](4,4.2){${\red G_\mathrm{rt}}$}
	\rput[c](2,2){\begin{tabular}{c}
	{\Large $\E8[\times\E8]$}\end{tabular}}
	\pnode(2.5,1.2){bulk}
\end{pspicture}
}}\right\}\xrightarrow[\text{effective theory}]{\text{`low-energy'}}
\left\{\begin{array}{l}
\text{{\purple \large{standard}}}\\
\text{{\purple \large{model}}}\\
\text{as an}\\
\text{intersection}\\
\text{of }{\red G_\mathrm{rb}}, {\red G_\mathrm{rt}}, {\red G_\mathrm{lt}}\\
\&\,{\blue \SO{10}}\\
\text{in }\E8[\times\E8]\end{array}
\right.$
\end{center}
\ncline[linewidth=1mm,linecolor=blue]{->}{gen}{SO10}
\ncarc[arcangle=-15]{->}{Higgs}{bulk}
\caption{Local grand unification.}
\label{fig:LocalGrandUnification}
\end{figure}

The key idea
of `local grand unification' \cite{Buchmuller:2005jr,Buchmuller:2005sh,Buchmuller:2006ik}, 
 which has been  the guiding principle in the MSSM `MiniLandscape' study
\cite{Lebedev:2006kn,Lebedev:2006tr,Lebedev:2007hv}, is that the structure of the
MSSM  can be explained in orbifolds with non-trivial gauge group
topography as follows (cf.\ figure \ref{fig:LocalGrandUnification}): 
\begin{itemize}
 \item  (most) matter fields  come from $\boldsymbol{16}$-plets living at the fixed
 points with \SO{10} symmetry;
 \item   $G_\mathrm{SM}\subset \SO{10} $   is the intersection of this `local
 \SO{10}' with other local gauge groups in $\E8\times \E8$;
 \item Higgs fields live (mostly)  in the bulk and appear  as split
 multiplets.  
\end{itemize}

\SO{10} local grand unification works only under certain conditions: one needs
fixed points  with \SO{10} symmetry admitting a massless $\boldsymbol{16}$-plet.
For example, \Z3 and $\Z2\times\Z2$  do not fall into this category.  The
simplest possibility would be a \Z4 orbifold; however it turns out that there it
is difficult to get three generations. The next-to-simplest possibility is then
a \Z6 orbifold. While the \Z6-I orbifold does not allow for a  local \SO{10} and
$G_\mathrm{SM}$ in 4D  at the same time (since there is only one Wilson line,
see \cite{Kobayashi:1990mi}), it turns out that \Z6-II, utilized by KRZ,
provides an excellent framework for the realization of local grand unification.
It is also interesting that \Z6-II is favored for other reasons
\cite{Ibanez:1992hc}.\footnote{Note that the nomenclature of \cite{Ibanez:1992hc} differs
from ours: what is called \Z6-I corresponds to \Z6-II in these notes.}

The most obvious option to get 3 generations  is to set the Wilson line in the \SU3 torus to
zero and use non-trivial Wilson lines in the \SO{4} torus. This guarantees triplication of 
families \cite{Buchmuller:2004hv}. 
However, it turns out that all such
models in \Z6-II  suffer from the presence of chiral exotics
\cite{Buchmuller:2006ik}. Similar considerations apply to all orbifolds up
to order 6 inclusive. 
One is therefore led to consider models
with 2 localized families where the third generation comes from the untwisted or
higher twisted sectors. This implies, in particular, that the theory $requires$
the first two and the third families to be fundamentally different. Needless to say,
this  result  does not go against the data.

\subsection{MSSM from heterotic orbifolds}
\label{sec:BHLR}

The strategy of starting with two localized generations and requiring 
 the third generation  to come from somewhere else turns out to be  successful. 
The first heterotic MSSM  with this structure has been presented in
\cite{Buchmuller:2005jr,Buchmuller:2006ik}. In this model, upon compactification
one finds
\begin{equation}
 \text{gauge group}~=~\SU3_\mathrm{C}\times\SU2_\mathrm{L}\times\U1_Y
 \times[\SU4\times\SU2\times\U1^8]\;,
\end{equation}
and
\begin{equation}
 \text{spectrum}~=~3\times\text{generation}+\text{vector-like} \;,
\end{equation}
where `vector-likeness' is meant with respect to 
 $G_\mathrm{SM}$. The gauge group is broken further by requiring the 
 Fayet-Iliopoulos (FI) $D$-term \cite{Dine:1987xk}  to be
cancelled, which implies non-zero vevs for some of the scalars. 
There are numerous possibilities to cancel the FI term while
maintaining vanishing $F$-terms, corresponding to a large number of
supersymmetric vacua. Most vacua are not phenomenologically interesting. What
is, however, important in the model of 
\cite{Buchmuller:2005jr,Buchmuller:2006ik} is that it admits supersymmetric
MSSM vacua where all the exotics become  massive. To establish the decoupling of exotics,
one has to prove that there are configurations where
\begin{itemize}
 \item couplings of the form
 \[ x_i\,\bar x_j\, \prod_\alpha s_{i_\alpha}\;,\]
 where $x_i$ and $\bar x_j$ denote the exotics and $s_{i_\alpha}$ are standard
 model singlets, exist such that the exotics' mass matrices have full rank;
 \item the $s_{i_\alpha}$ vevs are consistent with supersymmetry, i.e.\
  $F$- and $D$-terms are zero.
\end{itemize}
To identify such configurations  in practice is quite cumbersome (the interested reader is referred to
\cite{Buchmuller:2006ik,Lebedev:2007hv} for details). We note that
giving supersymmetric vevs to the singlets has further benefits such as
breaking the  unwanted gauge factors (rank reduction), and the generation of
effective Yukawa couplings, originating from higher-dimensional superpotential
terms. 

The main features of the MSSM vacua of the \Z6-II model 
\cite{Buchmuller:2005jr,Buchmuller:2006ik} are:
\begin{dingautolist}{"0CA}
 \item exact MSSM spectrum, i.e.\ 
 \[3\times \text{generation}+\text{Higgs}+\text{nothing}\;,\]
 where `nothing' means a hidden sector;
 \item gauge group is $G_\mathrm{SM}$ (up to a hidden sector), with hypercharge being that of
the usual \SU5;
 \item due to the above  features, the model is consistent with 
 MSSM gauge coupling unification;
 \item the top Yukawa coupling is order 1, while all other  couplings are 
small and generated {\it \`a la}  
 Froggatt-Nielsen  \cite{Froggatt:1978nt};
 \item there is an \SU4 group in the hidden sector  whose gauginos condense
\cite{Ferrara:1982qs,Nilles:1982ik,Derendinger:1985kk,Dine:1985rz} 
and can induce  TeV soft masses  
\end{dingautolist}
The model can also be viewed from the 6D orbifold GUT perspective 
which makes discussion of certain issues more tractable \cite{Buchmuller:2006ik,Buchmuller:2007qf}.

Of course, this is not the only string construction with realistic features. 
For instance, in Calabi-Yau (CY) compactifications it is also possible to obtain
models with the SM gauge group and  matter content up to vector-like exotics 
\cite{Pokorski:1998hr}.
Free fermionic constructions also provide  models with
realistic features \cite{Cleaver:1998sa,Cleaver:2007ek}.\footnote{See however \cite{Chaudhuri:1995ve}.} 
A three generation $Z'$ model based on the gauge group $G_\mathrm{SM}\times\U1_{B-L}$
has been constructed in
\cite{Braun:2005nv,Braun:2006ae}.\footnote{In this model,  $B-L$ cannot be
broken without breaking supersymmetry. Whether a viable model can be constructed  
is not yet clear.}
A  promising  model on a Calabi-Yau manifold   has also been reported in \cite{Bouchard:2005ag}. It has the
exact MSSM spectrum directly after compactification, and enjoys non-trivial Yukawa
couplings \cite{Bouchard:2006dn}. 
GUT-like  constructions, leading to flipped \SU5 models at the
compactification scale, have been discussed in
\cite{Blumenhagen:2006wj,Blumenhagen:2006ux}. 
Coming back to orbifold models,  \Z{12}-I  also  hosts  interesting models, either with
the flipped \SU5 \cite{Kim:2006hv,Kim:2006zw} or the SM gauge group \cite{Kim:2007mt}.
The   \Z3 orbifold, that has been under scrutiny for a long time,  
has been revived in the context 
of  multi-Higgs extensions of the MSSM  \cite{Munoz:2007sf,Escudero:2007db}.
More complicated orbifolds with non-factorizable tori and/or torsion are being 
actively studied \cite{Forste:2006wq,Takahashi:2007qc,
Ploger:2007iq}.

\subsection{Orbifold GUT limits}
\label{sec:OrbifoldGUTLimits}

Having obtained string models with a simple geometric interpretation and
realistic features, one might study orbifold GUT limits, which correspond to
anisotropic compactifications where some radii are significantly larger than the
others. This then leads to a `volume factor' which can explain the differences
between the 4D Planck scale, the string scale and the GUT scale
$M_\mathrm{GUT}$. $M_\mathrm{GUT}$ is, in this picture, essentially the inverse
of the largest compactification radius, while other the radii are required to be
smaller in order to stay in a regime where the description in terms of the
weakly coupled heterotic strings can be applied
\cite{Witten:1996mzF3,Hebecker:2004ce}.  Various orbifold GUT limits have been
derived in \cite{Kobayashi:2004ya,Buchmuller:2004hv,Buchmuller:2006ik}. (In
order to obtain a simple field-theoretic interpretation of the projection
conditions it is convenient to transform the shift and Wilson lines to a
particular form; see appendix A of \cite{Ploger:2007iq}.) Amazingly, it turns
out that in many cases the orbifold GUT limits are consistent with gauge
coupling unification \cite{Buchmuller:2004hv,Buchmuller:2006ik} in the following
sense: the standard model gauge factors get combined into a simple bulk gauge
group factor, such that at least the dominant threshold corrections (see
\cite{Dixon:1990pc,Mayr:1993mq,Stieberger:1998yi} for the string calculations)
are universal and do not spoil unification (see the discussion above figure
\ref{fig:OGL} for an example).

\subsection{Heterotic MiniLandscape}

Having seen a particular example of the MSSM from the  \Z6-II orbifold
(subsection \ref{sec:BHLR}), it is imperative to ask whether more  comparable
models exist in this construction and whether they are frequent. This question
has been answered affirmatively in
\cite{Lebedev:2006kn,Lebedev:2006tr,Lebedev:2007hv}, where the \Z6-II models
with two localized $\boldsymbol{16}$-plets of \SO{10} have been analyzed
systematically. 

In the \Z6-II orbifold, there are two  gauge  shifts that produce a local
\SO{10} GUT with  $\boldsymbol{16}$-plets (cf.\ \cite{Katsuki:1989cs}),
\begin{eqnarray}
 V^{ \SO{10},1} &= &
 \left(\tfrac{1}{3},\,\tfrac{1}{2},\,\tfrac{1}{2},\,0,\,0,\,0,\,0,\,0\right)\,\left(\tfrac{1}{3},\,0,\,0,\,0,\,0,\,0,\,0,\,0\right)
 \;,
 \nonumber \\
 V^{ \SO{10},2 }& = &
 \left(\tfrac{1}{3},\,\tfrac{1}{3},\,\tfrac{1}{3},\,0,\,0,\,0,\,0,\,0\right)\,\left(\tfrac{1}{6},\,\tfrac{1}{6},\,0,\,0,\,0,\,0,\,0,\,0\right)
 \;. \label{eq:so10shifts}
\end{eqnarray}
For each of these shifts, we follow the steps:
\begin{dingautolist}{"0C0}
 \item Generate Wilson lines $W_3$ and $W_2$.
 \item Identify ``inequivalent'' models.
 \item Select models with $G_\mathrm{SM} \subset \SU5 \subset \SO{10}$.
 \item Select models with three net
 $(\boldsymbol{3},\boldsymbol{2})$.
 \item Select models with non-anomalous $\U1_{Y} \subset \SU5$.
 \item Select models with net 3 SM families + Higgses + vector-like.
\end{dingautolist}
It turns out that in these models almost $1\,\%$ has the MSSM spectrum plus
vector-like exotics (table \ref{tab:Statistics}).

\begin{table*}[h!]
\centerline{
\begin{tabular}{|l||l|l|}
\hline
 criterion & $V^{\SO{10},1}$ & $V^{\SO{10},2}$ \\
\hline
&&\\[-0.3cm]
 \ding{"0C1}  inequivalent models with 2 Wilson lines
  &$22,000$ & $7,800$   \\[0.2cm]
  \ding{"0C2} SM gauge group $\subset$ SU(5) $\subset$ SO(10)
  ({\rm or}~\E6)
  &3563 &1163 \\[0.2cm]
  \ding{"0C3} 3 net $(\boldsymbol{3},\boldsymbol{2})$
  &1170 &492 \\[0.2cm]
  \ding{"0C4} non-anomalous $\U1_{Y}\subset \SU5 $
  &528 &234 \\[0.2cm]
  \ding{"0C5}  spectrum $=$ 3 generations $+$ vector-like
  &128 &90 
  \\
\hline
\end{tabular}
}
\caption{Statistics of \Z6-II orbifolds based on the SO(10) shifts
\eqref{eq:so10shifts} with two Wilson lines. \label{tab:Statistics} }
\end{table*}

It is instructive to compare these results to other MSSM searches in the
literature. In certain types of intersecting D-brane models, it was found that
the probability of obtaining the SM gauge group and three generations of quarks
and leptons, while allowing for chiral exotics, is less than $10^{-9}$
\cite{Gmeiner:2005vz,Douglas:2006xy}. The criterion which comes closest to the
requirements imposed in \cite{Gmeiner:2005vz,Douglas:2006xy} is \ding{"0C3}.  We
find that within our sample the corresponding probability is 6\,\%. In
\cite{Dijkstra:2004cc,Anastasopoulos:2006da}, orientifolds of Gepner models were
scanned for chiral MSSM matter spectra, and it was found that the fraction of
such models is $4 \times 10^{-14}$. These constructions contain the MSSM matter
spectrum plus  vector-like exotics.   This is most similar to step \ding{"0C5}
in our analysis where we find 218 models out of a total of $3 \times 10^4$ or
0.7\,\%. In comparison, approximately 0.6\,\% of our models have the MSSM
spectrum at low energies with all vector-like exotics decoupling along $D$-flat
directions. Note also that, in all of our models, hypercharge is normalized as
in standard GUTs and thus consistent with gauge coupling unification. We learn
from this comparison that the heterotic orbifolds with local \SO{10} structure
are particularly ``fertile'' in producing the MSSM.

Having obtained a set of $\mathcal{O}(100)$ models with realistic features, one
can (and should) study their properties. An interesting question is whether
realistic features are correlated. In \cite{Lebedev:2006tr} the distribution of
the scales of gaugino condensation in the models with an exact MSSM spectrum was
studied. The scheme of gaugino condensation provides a
natural explanation of the hierarchy between the Planck and supersymmetry breaking
 (or electroweak) scales. Gravity mediated SUSY breaking predicts 
an approximate  relation
\begin{equation}
 m_{3/2}~\sim~\frac{\Lambda^3}{M_\mathrm{P}^2}\;,
\end{equation}
where the gravitino mass $m_{3/2}$ sets the scale for the MSSM soft masses and
$\Lambda$ is the scale of the hidden sector strong dynamics. 
The distribution of  $\Lambda$  in  our models    shows
that demanding realistic features in the  observable sector leads to
the preference for  gaugino condensation at an intermediate scale. 
Therefore these models  favour TeV-scale MSSM soft masses which  provides a top-down motivation for low-energy 
supersymmetry \cite{Lebedev:2006tr}.\footnote{%
We would also like to comment that in the context of K\"{a}hler stabilization
\cite{Binetruy:1996xj,Casas:1996zi,Gaillard:2007jr}, the hidden sector strong
dynamics allows to fix the dilaton, as well as the other moduli, in particular
the $T$-moduli \cite{Font:1990nt,Nilles:1990jv,Barreiro:1997rp}, which
parametrize the radii of the three tori. Of course, moduli stabilization is a
complicated issue, and there are many different possibilities (see e.g.\
\cite{deCarlos:1993da,Serone:2007sv}). In \cite{Font:1990nt,Nilles:1990jv} it
was found that the $T$-moduli get fixed at order one values (in Planckian
units), on the other hand, as briefly discussed in subsection
\ref{sec:OrbifoldGUTLimits}, it appears desirable to obtain anisotropic
compactifications. The results of
\cite{Font:1990nt,Nilles:1990jv,deCarlos:1993da} were obtained in the context of
\Z3 orbifolds without Wilson lines, where threshold corrections attain a
particularly simple form \cite{Dixon:1990pc,Mayr:1993mq,Stieberger:1998yi}. It
should be interesting to study whether in the \Z6-II orbifold
with discrete Wilson lines there are hierarchies between the $T$-moduli vevs,
leading to anisotropic compactifications.}

\subsection{A heterotic `benchmark model'}

Somewhat surprisingly, the models of the MiniLandscape possess even more
attractive features such as $R$-parity, the neutrino seesaw, etc.  To be specific,
let us focus on an explicit example: the vacuum of one of the   MiniLandscape
models, referred to as `1A' in \cite{Lebedev:2007hv}. We would, however, like to
emphasize that these features are shared by many other vacua of the
MiniLandscape models.

\paragraph{$\boldsymbol{R}$-parity.} As in grand unification, it is possible
to obtain $R$-parity as a \Z2 subgroup of  $\U1_{B-L}$. An important difference 
however is that $\U1_{B-L}$ is not embedded in \SO{10} and 
there can be SM singlets with even $B-L$ charge
(not related to $\overline{\boldsymbol{126}}$-plets of \SO{10}). 
This
facilitates the construction of MSSM vacua with an \emph{exact} $R$-parity
\cite{Lebedev:2006kn,Lebedev:2007hv}, $\Z2^R$. This is to be contrasted with
\cite{Buchmuller:2006ik} where $R$-parity was approximate, and to
\cite{Bouchard:2006dn} where $R$-parity exists only at the classical level
and  $Y_e$ and $Y_d$ vanish at the same level. 
The model also does not suffer
from the problem encountered in \cite{Kim:2007mt}, where it was found that one
can either decouple all exotics or have $R$-parity but not both.

\paragraph{$\boldsymbol{\mu}$-term.} In model 1A, there is a vector-like pair of
weak doublets, $\phi_u$ and $\phi_d$, from the (third) untwisted sector.  These 
correspond to the extra components of the 10D gauge bosons.\footnote{This
property is shared by all models based on shift $V^{\SO{10},1}$.} They arise
from an \SO{10} $\boldsymbol{10}$-plet contained in the adjoint of \E8. The pair
$\phi_u\,\phi_d$ is neutral w.r.t.\ the selection rules, i.e.\ whenever a
superpotential term $\psi_1^{n_1}\cdots \psi_r^{n_r}$ is allowed, also the term 
$\phi_u\,\phi_d\cdot\psi_1^{n_1}\cdots \psi_r^{n_r}$ is allowed. Further, it was
found \cite{Lebedev:2007hv} that (at least at a given order) the global SUSY
$F$-term equations are satisfied term by term,
\begin{equation}
 \frac{\partial\mathscr{M}}{\partial s_i}~=~0\;,
\end{equation}
where $\mathscr{M}$ denotes a monomial of SM singlets $s_i$
entering the superpotential. One finds that
also $\mathscr{M}$ vanishes. 
That is, in supersymmetric vacua where
the $F$-terms vanish term by term, the vev of the superpotential is zero. 
Further, the bilinear couplings involving $\phi_u$ or $\phi_d$ to any other
\SU2 doublet vanish because of the $G_\mathrm{SM}\times \Z2^R$ symmetries 
\cite{Lebedev:2007hv}. The  $\phi_u\,\phi_d$ mass
term can only be due to the above monomials $\mathscr{M}$. Since the latter
is zero in the vacuum, the $\mu$-term is zero and exactly one pair of Higgs doublets 
is massless (while the exotics are decoupled). 
 In other
words, demanding that
\begin{itemize}
 \item $G_\mathrm{SM}\times \Z2^R$ be unbroken and
 \item $F$-terms vanish term by term
\end{itemize}
in this model leads to supersymmetric vacua with a suppressed $\mu$-term.
When SUSY gets broken, the $\mu$-term of the order of the gravitino mass 
is generated.
 This constitutes a stringy solution to the MSSM $\mu$ problem.

\paragraph{Gauge-top unification.} The fact that the Higgs doublets stem from
the untwisted sector has further important consequences. The left- and right-handed
up-type quarks of the third generation are also untwisted and the corresponding interaction 
 $\phi_u\,\bar u\,q$ is allowed by string selection rules. 
Furthermore, its strength is given by the gauge coupling since it stems from 
supergauge interactions in 10D. Thus the top Yukawa is predicted to be
the same as the gauge coupling at the string scale.\footnote{Again, 
this property to a large extent  is shared by all models based on shift
$V^{\SO{10},1}$, in particular by the model of 
\cite{Buchmuller:2005jr,Buchmuller:2005sh}.}
The   other Yukawa
couplings appear at higher orders 
 and are therefore suppressed, similarly to the
Froggatt-Nielsen picture \cite{Froggatt:1978nt}. The idea to relate the
top Yukawa coupling to the gauge coupling is not new (see e.g.\
\cite{Kubo:1985up,Kubo:1994bj}), however, the fact that this happens automatically in
many models is  remarkable.

\paragraph{Neutrino masses and see-saw.} To discuss the neutrino masses in
string-derived models one has first to clarify what a right-handed neutrino is.
In supersymmetric MSSM vacua with  $R$-parity, this question is answered quite easily:
a right-handed  neutrino is an $R$-parity odd $G_\mathrm{SM}$ singlet. In the model discussed
so far, there are 49 such  neutrinos. Further, as discussed more generally in
\cite{Buchmuller:2007zd}, all ingredients of the see-saw \cite{Minkowski:1977sc}
are present:
\begin{itemize} 
 \item the right-handed neutrino mass matrix $M_\nu$ has full rank and
 \item neutrino Yukawa couplings $Y_\nu$ exist 
\end{itemize}
The resulting  effective  mass matrix for the light neutrinos,
\begin{equation}
 m_\nu~=~v_u^2\cdot Y_\nu^T\,M^{-1}\,Y_\nu\;,
\end{equation}
where $v_u$ denotes the vev of the up-type Higgs $\phi_u$, has full rank.
Thus, all light neutrinos have a small mass as a result of the seesaw
mechanism.  Due to the large number of neutrinos, the effective neutrino mass
operator has  many contributions such that neutrino masses are  enhanced
compared to  the naive estimate $m_\nu^\mathrm{naive}\sim v_u^2/M_\mathrm{GUT}$.
Let us also mention that meanwhile the many neutrino scenario has been analyzed
in some detail. It has been found that the presence of many right-handed
neutrinos helps ameliorate  the tension between leptogenesis and supersymmetry
\cite{Eisele:2007ws,Ellis:2007wz}. Furthermore, this scenario has important
implications for supersymmetric lepton flavor violation  and electric dipole
moments \cite{Ellis:2007wz}.

\paragraph{Proton stability.} Because of the exact $R$-parity, dimension four
proton decay operators are absent. However,  both
$q\,q\,q\,\ell$ and $\bar u\,\bar u\,\bar d\,\bar e$  appear at order 6 in the SM
singlets, and  are also generated by integrating out the heavy exotics.
This leads to  effective dimension five  operators mediating proton decay, 
the usual problem  of  4D GUTs
\cite{Dermisek:2000hr}.
Forbidding such operators is likely to  require further (perhaps discrete) symmetries
\cite{Mohapatra:2007vd},  which
 is currently under investigation.

\paragraph{Non-Abelian discrete flavor symmetries.} The models of the
MiniLandscape exhibit a non-Abelian discrete flavor symmetry, $D_4$, for the two
light generations \cite{Kobayashi:2004ya,Kobayashi:2006wq}. This symmetry is
only exact when the SM singlets have zero vevs, and  is broken in realistic vacua. 
If the breaking is small, $D_4$ can be relevant to the (supersymmetric)
flavor structure (cf.~\cite{Ko:2007dz}).

\begin{figure}[!h!]
\begin{center}
\CenterObject{\psset{unit=1.2cm}
\begin{pspicture}(-2,-0.3)(6.5,4.5)
\pscustom[slopesteps=200,fillstyle=ccslope,slopebegin=white,slopeend=gray,slopecenter=0.7 0.7]{
	\pscurve[liftpen=1](0,0)(2,0.2)(4,0)
	\pscurve[liftpen=1](4,0)(3.7,2)(4,4)
	\pscurve[liftpen=1](4,4)(2,3.8)(0,4)
	\pscurve[liftpen=1](0,4)(0.3,2)(0,0)
	}
	\pscurve[linewidth=0.5mm](0.8,0.28)(2,0.4)(3.2,0.28)
	\pscurve[linewidth=0.5mm](1.2,0.5)(2,0.56)(2.8,0.5)
	\pscurve[linewidth=0.5mm](0.8,3.72)(2,3.6)(3.2,3.72)
	\pscurve[linewidth=0.5mm](1.2,3.5)(2,3.44)(2.8,3.5)
	\pscurve[linewidth=0.5mm](0.36,0.8)(0.48,2)(0.36,3.2)
	\pscurve[linewidth=0.5mm](0.6,1.2)(0.68,2)(0.6,2.8)
	\pscurve[linewidth=0.5mm](3.64,0.8)(3.52,2)(3.64,3.2)
	\pscurve[linewidth=0.5mm](3.4,1.2)(3.32,2)(3.4,2.8)
	\psdot*[dotstyle=o,dotsize=3mm,linecolor=gray](0,0)
 	\psdot*[dotstyle=square*,dotsize=2mm,linecolor=gray](0,0)
	\psdot*[dotstyle=o,dotsize=3mm,linecolor=gray](4,0)
	\psdot*[dotstyle=square*,dotsize=2mm,linecolor=gray](4,0)
	\psdot*[dotstyle=o,dotsize=3mm,linecolor=gray](4,4)
	\psdot*[dotstyle=square*,dotsize=2mm,linecolor=gray](4,4)
	\psdot*[dotstyle=o,dotsize=3mm,linecolor=gray](0,4)
	\psdot*[dotstyle=square*,dotsize=2mm,linecolor=gray](0,4)
	\rput[c](2,2){\begin{tabular}{c}
	{\Huge $\SU{6}$}\\[0.3cm]
	{\Large \blue $3^\mathrm{rd}$ family}\end{tabular}}
	\pnode(2.5,1.2){bulk}
	\psdot*[dotstyle=o,dotsize=3mm,linecolor=blue](0,0)
 	\psdot*[dotstyle=square*,dotsize=2mm,linecolor=blue](0,0)
	\rput[t](0,-0.3){$\SU5\subset{\blue \SO{10}}$}
	\rput[r](-0.3,0.3){$\boldsymbol{10}+\overline{\boldsymbol{5}}+\boldsymbol{1}=
	{\blue \boldsymbol{16}}$}
	\psdot*[dotstyle=o,dotsize=3mm,linecolor=blue](0,4)
 	\psdot*[dotstyle=square*,dotsize=2mm,linecolor=blue](0,4)
	\rput[b](0,4.3){$\SU5\subset{\blue \SO{10}}$}
	\rput[r](-0.3,3.7){$\boldsymbol{10}+\overline{\boldsymbol{5}}+\boldsymbol{1}=
	{\blue \boldsymbol{16}}$}
	\rput[t](4,-0.3){$\SU4\times\SU2_\mathrm{L}$}
	\rput[b](4,4.3){$\SU4\times\SU2_\mathrm{L}$}	
\end{pspicture}}
\end{center}
\caption{Orbifold GUT limit of model 1A of \cite{Lebedev:2007hv}.}	
\label{fig:OGL}	
\end{figure}
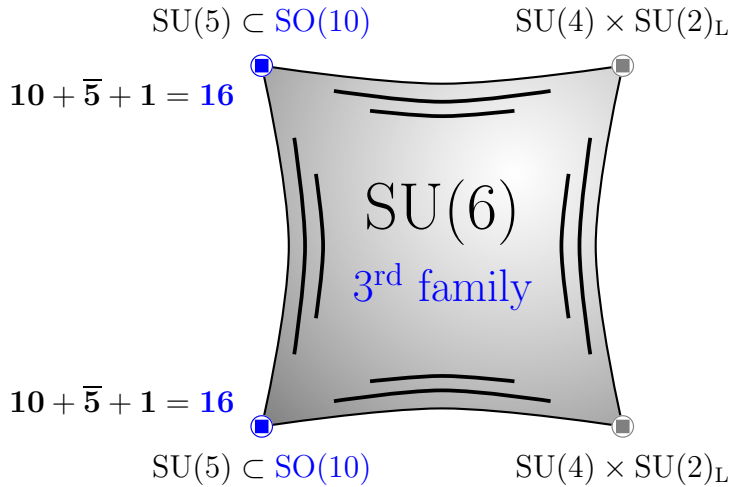

\paragraph{Orbifold GUT limits.} Last but not least, let us discuss orbifold GUT
limits (cf.\ subsection~\ref{sec:OrbifoldGUTLimits}) of this model. Consider the
case where the radii of the \SO4 torus, $R_{\SO4}$, are significantly larger
than the radii of the \SU3 and \G2 tori, $R_\mathrm{other}$. In the energy
range  $R_{\SO4}^{-1}<E<R_\mathrm{other}^{-1}$ the model can be described by an
effective 6D theory with an \SU6 bulk symmetry that gets broken to $\SU5$ and
$\SU4\times\SU2_\mathrm{L}$ at the two left or the two right fixed points in
figure \ref{fig:OGL}, respectively (we are ignoring the second \E8 and \U1
factors). The intersection of \SU5 and $\SU4\times\SU2_\mathrm{L}$ in \SU6 is
the SM gauge group $G_\mathrm{SM}$. Since all standard model gauge group factors
are contained in the bulk \SU6, threshold corrections are universal, and the
model is consistent with MSSM gauge coupling unification. Two of the three SM
families reside at two equivalent \SU5 fixed points while the third generation
comes from the bulk (see figure~\ref{fig:OGL}). In order to understand why
matter on the \SU5 fixed points combines to complete generations, i.e.\
$\boldsymbol{16}$-plets of \SO{10}, one has to zoom into the fixed points and to
resolve the underlying \SO{10} structure. By making the vertical direction in
figure~\ref{fig:OGL} small as well, one arrives at a setting that strongly
resembles the Kawamura model \cite{Kawamura:1999nj,Kawamura:2000ev}, except for
the fact that the third family lives on the interval rather than on the boundary
(cf.\ the analogous discussion in Ref.~\cite{Buchmuller:2007qf}). 

%\clearpage

\section{Summary}

In summary, the scheme of local grand unification provides an attractive
framework for connecting observations to fundamental physics. It shares many
attractive features with conventional  4D  GUTs while avoiding their most
problematic aspects such as  the doublet-triplet splitting problem and
unrealistic fermion mass relations. In the context of string theory, the concept
of local grand unification facilitates  construction of phenomenologically
attractive models. These are automatically anomaly-free, UV complete and also
incorporate gravity. They also exhibit some surprising features   including  a 
novel stringy solution to the MSSM $\mu$ problem.  Clearly, it is imperative to
seek  a more comprehensive understanding  of these constructions.

\section*{Acknowledgements}

I'm indebted to the organizers of the Summer Institute for the wonderful meeting
and the nice hikes. I would like to thank W.~Buchm\"uller, K.~Hamaguchi, 
T.~Kobayashi, O.~Lebedev, H.P.~Nilles, F.~Pl\"oger, S.~Raby, S.~Ramos-S\'anchez
and P.~Vaudrevange for very pleasant collaborations, and M.-T.~Eisele,
S.~Ramos-S\'anchez, K.~Schmidt-Hoberg and, in particular, O.~Lebedev for
comments on the manuscript.  Further thanks go to the Aspen Center for Physics,
where parts of these notes were written, for hospitality and support. This
research was supported by the DFG cluster of excellence Origin and Structure of
the Universe  and by the SFB-Transregio 27 "Neutrinos and Beyond".

\bibliography{Orbifold}
\bibliographystyle{NewArXiv}

\end{document}